# Design Topics for Superconducting RF Cavities and Ancillaries

*H. Padamsee*[1]
Cornell University, CLASSE, Ithaca, New York

**Abstract**
RF superconductivity has become a major subfield of accelerator science. There has been an explosion in the number of accelerator applications and in the number of laboratories engaged. The first lecture at this meeting of the CAS presented a review of fundamental design principles to develop cavity geometries to accelerate velocity-of-light particles ($\beta = v/c \sim 1$), moving on to the corresponding design principles for medium-velocity (medium-$\beta$) and low-velocity (low-$\beta$) structures. The lecture included mechanical design topics. The second lecture dealt with input couplers, higher-order mode extraction couplers with absorbers, and tuners of both the slow and fast varieties.

## 1     General advantages of SRF cavities

There are two textbooks covering the major topics [1, 2]. Many review articles [3–13] are now available covering the state of the art. There have been 16 international workshops on RF superconductivity. The proceedings from these workshops carry a detailed and comprehensive coverage of the substantial work going on, with many excellent tutorials on special subjects. These proceedings are available in electronic form on the JACOW website (http://jacow.org/). Basics covered in [1] are not repeated here, although essentials are summarized. Selective examples presented here are for illustrative purposes only.

Superconducting Radio Frequency (SRF) cavities excel in applications requiring Continuous Wave (CW) or long-pulse accelerating fields above a few million volts per meter ($MV \cdot m^{-1}$). We often refer to the accelerating field as the 'gradient'. Since the ohmic (power) loss in the walls of a cavity increases as the square of the accelerating voltage, copper cavities become uneconomical when the demand for high CW voltage grows with particle energy. A similar situation prevails in applications that demand long RF pulse length, or high RF duty factor. Here superconductivity brings immense benefits. The surface resistance of a superconducting cavity is many orders of magnitude less than that of copper. Hence the intrinsic quality factor ($Q_0$) of a superconducting cavity is usually in the $10^9$ to $10^{10}$ range. Characterizing the wall losses, $Q_0$ is a convenient parameter for the number of oscillations it takes the stored energy in a cavity to dissipate to zero ($Q_0$ is often abbreviated here as $Q$). After accounting for the refrigerator power needed to provide the liquid helium operating temperature, a net gain factor of several hundred remains in the overall operating power for superconducting cavities over copper cavities. This gain provides many advantages.

Copper cavities are limited to gradients near $1\,MV \cdot m^{-1}$ in CW and long-pulse (duty factor > 1%) operation because the capital cost of the RF power and the ac-power related operating cost become prohibitive. For example, several $MW \cdot m^{-1}$ of RF power are required to operate a copper cavity at $5\,MV \cdot m^{-1}$. There are also practical limits to dissipating such high power in the walls of a copper cavity. The surface temperature becomes excessive, causing vacuum degradation, stresses, and metal fatigue due to thermal expansion. On the other hand, copper cavities offer much higher accelerating fields (~100 MV·m$^{-1}$) for short pulse ($\mu$s) and low duty factor (<0.1%) applications. For such applications it is still necessary to provide abundant peak RF power (e.g. 100 MW·m$^{-1}$), and to prevent or withstand the aftermath of intense voltage breakdown in order to reach the very high fields.

---
[1] hsp3@cornell.edu

There is another important advantage that SRF cavities bring to accelerators. The presence of accelerating structures has a disruptive effect on the beam, limiting the quality of the beam in aspects such as energy spread, beam halo, or even the maximum current. Because of their capability to provide higher voltage, SRF systems can be shorter, and thereby impose less disruption. Due to their high ohmic losses, the geometry of copper cavities must be optimized to provide a high electric field on axis for a given wall dissipation. This requirement tends to push the beam aperture to small values, which disrupts beam quality. By virtue of low wall losses, it is affordable to design an SRF cavity to have a large beam hole, reduce beam disruption, and provide high-quality beams for physics research.

For low-velocity, heavy ion accelerators [7, 8], which must be capable of accelerating a variety of ion species and charge states, a major advantage of superconducting resonators is that a CW high voltage can be obtained in a short structure. The desired linac can be formed as an array of independently phased resonators, making it possible to vary the velocity profile of the machine. An independently phased array provides a high degree of operational flexibility and tolerates variations in the performance of individual cavities. Superconducting boosters show excellent transverse and longitudinal phase space properties, and excel in beam transmission and timing characteristics. Because of their intrinsic modularity, there is also the flexibility to increase the output energy by adding higher velocity sections at the output, or to extend the mass range by adding lower velocity resonators at the input.

## 2 Figures of merit for cavity performance

The main figures of merit for an accelerating structure are defined and discussed in [1]. These are: RF frequency, accelerating voltage ($V_c$), accelerating field ($E_{acc}$), peak surface electric field ($E_{pk}$), peak surface magnetic field ($H_{pk}$), surface resistance ($R_s$), geometry factor ($G$), dissipated power ($P_c$), stored energy ($U$), $Q$ value, geometric shunt impedance ($R_{sh}/Q_0$, often mentioned as $R/Q$ for short), cell-to-cell coupling for multi-cell structures, Lorentz-Force (LF) detuning coefficient, input power required for beam power ($P_b$), coupling strength of input coupler ($Q_{ext}$), higher-order mode frequencies, and shunt impedances.

We present an in-depth discussion of several important figures of merit. The cavity accelerating voltage $V_c$ is the ratio of the maximum energy gain that a particle moving along the cavity axis can achieve to the charge of that particle. The accelerating gradient is defined as the ratio of the accelerating voltage per cell $V_c$ to the cell length. As the optimal length of the cavity cells is typically $\beta\lambda/2$, the accelerating gradient is

$$E_{acc} = \frac{V_c}{\beta\lambda/2} .$$

The RF power dissipation in a cavity wall is characterized by the quality factor $Q_0$, which tells us how many RF cycles (multiplied by $2\pi$) are required to dissipate the energy U stored in the cavity:

$$Q_0 = \frac{\omega_0 U}{P_c} = \frac{\omega_0 \mu \int_V |\mathbf{H}(\mathbf{r})|^2 \, dV}{\oint_A R_s |\mathbf{H}(\mathbf{r})|^2 \, dA} ,$$

where $P_c$ is the RF power dissipated in the cavity. The RF magnetic field $H(r)$ for the excited eigenmode with angular frequency $\omega_0 = 2\pi f_0$ is integrated over the cavity volume $V$ and surface $A$. The surface resistivity $R_s$ quantifies the RF power and depends only on the frequency and intrinsic material properties. It remains the only term in the formula that is material dependent, making it convenient to write the quality factor as

$$Q_0 = \frac{G}{\langle R_s \rangle}.$$

where $G$ is the geometry factor. The surface resistivity is a function of the RF magnetic field and may therefore vary along the cavity wall. It must be averaged over the cavity surface. The geometry factor $G$ is determined only by the shape of the cavity, and hence is useful for comparing cavities with different shapes. The cavity's shunt impedance $R_{sh}$ relates the dissipated power $P_c$ and the accelerating voltage:

$$P_c = \frac{V_c^2}{R_{sh}}.$$

A related quantity is the geometric shunt impedance $R_{sh}/Q_0$, or simply $R/Q$, which depends only on the cavity's shape. Two key figures of merit are the ratios of the peak surface electric and magnetic fields to the accelerating gradient, $E_{pk}/E_{acc}$ and $B_{pk}/E_{acc}$. A high surface electric field can cause field emission of electrons, thus degrading performance. A high surface magnetic field may limit the cavity's ultimate gradient performance by breakdown of superconductivity, also called quench.

## 2.1 Design choices

Taking into consideration the above figures of merit, some of the main choices that need to be made for structure design are: cavity frequency, cell shape, number of cells, beam aperture, operating gradient, operating temperature, input coupler, and Higher-Order Mode (HOM) coupler types. Two classes of considerations govern the choices: the particular accelerator application and the superconducting RF properties. Typical accelerator aspects are: the velocity of the particle(s) under acceleration, the desired voltage, the duty factor of accelerator operation, and the beam current or beam power. Other properties of the beam, such as bunch length, also play a role, as these influence the longitudinal and transverse wakefields, along with higher-order mode impedances. Typical superconducting properties influencing design choices are the microwave surface resistance at the chosen frequency, and the peak surface electric and magnetic fields at the design accelerating field. These properties set the operating field levels, the RF power required, as well as the ac operating power, together with the operating temperature. Mechanical properties also play a role to ensure stability under atmospheric loading and temperature differentials, to minimize LF detuning, and to keep microphonics detuning under control. Finally, input and output power coupling issues interact with cavity design.

Electromagnetic software packages for modelling accelerating cavities and couplers have been in existence for decades, first in 2D and later in 3D. Direct simulations of the entire cavity with input and HOM couplers have been carried out.

In general there are many trade-offs between competing requirements. For example, the higher the power capability of the input coupler, the larger the allowed number of cells per structure. But the difficulties of handling long structures set an upper limit on the number of cells. A large number of cells will also increase the probability of some HOMs remaining trapped inside the structure. A large beam aperture will improve the propagation of HOMs out of the structure, but will increase the peak surface electric and magnetic fields.

# 3 Classification of structures

There are three major classes of superconducting accelerating structures: high-, medium-, and low-$\beta$. Figure 1 shows some practical geometries for each type depending on the velocity of the particles, spanning the full velocity range of particles [7].

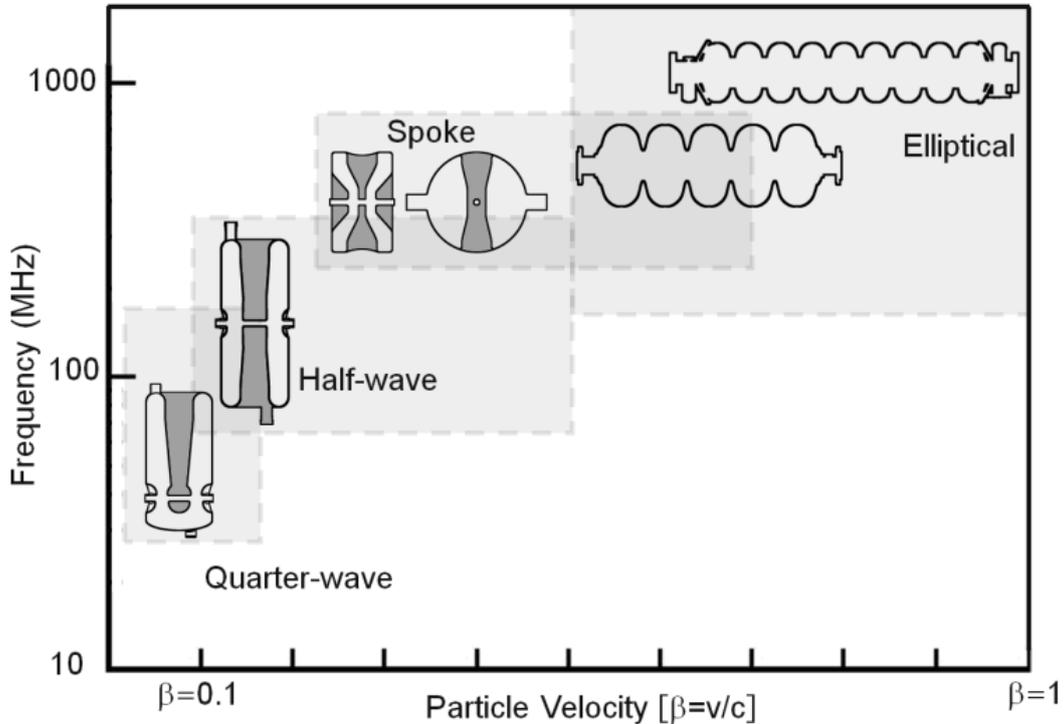

**Fig. 1:** Superconducting cavities spanning the full range of $\beta$ (reproduced from [7])

The high-$\beta$ structure, based on the $TM_{010}$ resonant cavity, is for acceleration of electrons, positrons, or high-energy protons with $\beta \sim 1$. The cavity gap length is usually $\beta\lambda/2$, where $\lambda$ is the wavelength corresponding to the frequency choice for the accelerating structure. Medium-velocity structures with $\beta$ between 0.2 and 0.7 are used for protons with energies less than 1 GeV as well as for ions. At the higher-$\beta$ end, these resonators are 'foreshortened' speed-of-light structures with longitudinal dimensions scaled by $\beta$. Near $\beta = 0.5$ spoke resonators with single or multi-gaps become popular. Spoke resonators operate in a TEM mode, and are so classified. The overlap between foreshortened elliptical and spoke structures near $\beta = 0.5$ involves several trade-offs, which we will discuss. Elliptical shape cells for $\beta < 0.5$ become mechanically unstable as the accelerating gap shortens and cavity walls become nearly vertical. The choice of a low RF frequency, favoured for ion and proton applications, also makes the elliptical cells very large, aggravating the structural weakness.

## 3.1 High-$\beta$ cavities

A typical high-$\beta$ accelerating structure consists of a chain of coupled cells operating in the $TM_{010}$ mode, where the phase of the instantaneous electric field in adjacent cells is shifted by $\pi$ to preserve acceleration as a charged particle traverses each cell in half an RF period. Figure 2 shows a nine-cell accelerating structure [14, 15] developed by the TESLA collaboration and used at FLASH (formerly the Tesla Test Facility, TTF). The beam enters and exits the structure via the beam tubes. Input coupler devices attached to ports on the beam tubes bring RF power into the cavity to establish the field and deliver beam power. Higher-order mode (HOM) couplers extract and damp the HOMs excited by the beam, and smaller ports carry pick-up probes to sample the cavity field for regulation and monitoring. The TESLA cavity will be used in the European X-ray Free Electron Laser (XFEL), and remains a strong candidate for the International Linear Collider (ILC).

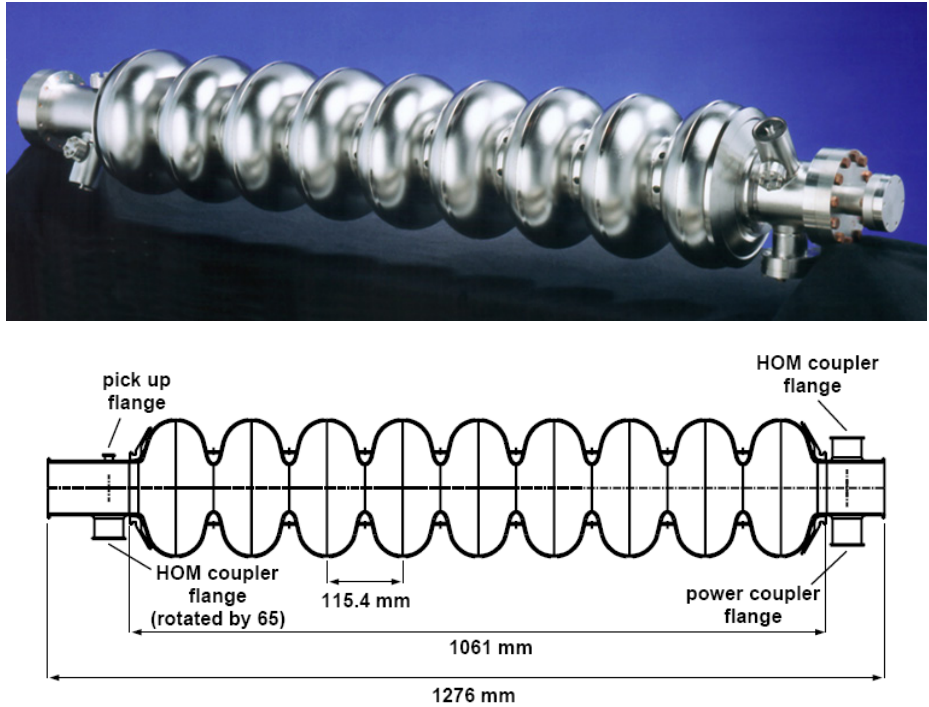

**Fig. 2:** Top: photograph of a nine-cell TESLA accelerating structure with one input power coupling port at one end and one HOM coupler at each end. Bottom: lay-out of the components for the nine-cell TESLA-style structure.

Single-cell cavities generally used for SRF R&D also find accelerator application, as, for example, in high current ring colliders, such as CESR, KEK-B, as well as many storage ring light sources. Figure 3 shows the single-cell CESR and KEK-B cavities [16–18].

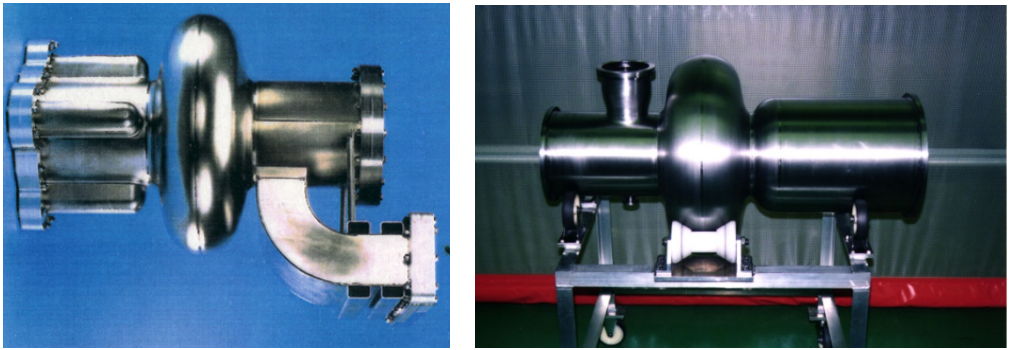

**Fig. 3:** Left: single cell 500 MHz cavity for CESR with waveguide input coupler and fluted beam tube on one side to remove the first dipole HOMs. The cell length is about 28 cm, slightly shorter than $\lambda/2$, to optimize the cell shunt impedance. Right: single-cell 508 MHz cavity for KEK-B with coaxial input coupler port and large beam pipe on one side for propagation of HOMs [18].

Most $\beta = 1$ structures are now based on the elliptical cavity. The elliptical cell shape [19] emerged from the more rounded 'spherical' shape [20], which was first developed to eliminate multipacting. The tilt of the elliptical cell also increases the stiffness against mechanical deformations and provides a better geometry for acid draining and water rinsing.

### 3.2 Multicell cavities

A multicell cavity is a structure with multiple resonators (cells) electromagnetically coupled together. For each mode of a single-cell cavity there are $N$ modes of excitation for an $N$-cell structure. For the fundamental $TM_{010}$ mode, there are $N$ $TM_{010}$-like modes. The accelerating mode is the one which provides an equal voltage kick to charged particles passing each cell. In this mode the fields in neighbouring cells are $\pi$ rad out of phase with each other. Thus, a particle moving at near the speed of light crosses each cell in half the RF period. The frequencies $f_m$ of the modes can be derived from a circuit model (Fig. 4) of the $N$-cell cavity as a series of coupled $LC$ resonators, where $L$ and $C$ are the characteristic inductance and capacitance for each cell.

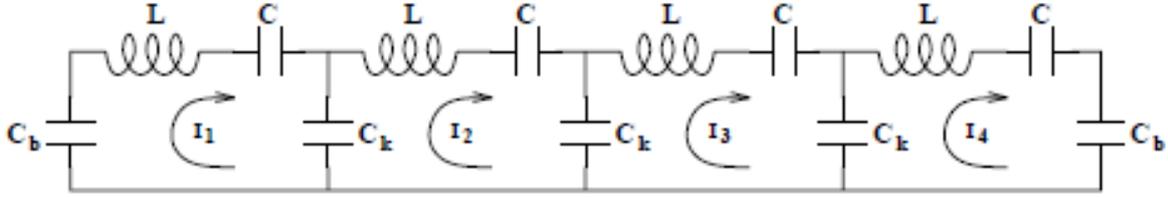

**Fig. 4:** Lumped circuit model for a multicell cavity

Here, $L$ and $C$ are related to the single-cell frequency $\omega_0$ and the shunt impedance $R_{sh}/Q_0$ via

$$\omega_0 = \sqrt{1/LC}, \quad \frac{R_{sh}}{Q_0} = 2\sqrt{L/C}.$$

The cell-to-cell coupling is related to the inter-cell capacitance $C_k$ via $k = C/C_k$, and $C_b$ is the capacitance representing the beam holes. The solution for the coupled Kirchhoff current and voltage equations yields the dispersion relation for mode frequencies $f_m$ for mode $m$ via

$$\Omega^{(m)} = \left(\frac{f_m}{f_0}\right) = 1 + 2k\left[1 - \cos\left(\frac{m\pi}{N}\right)\right].$$

As shown in Fig. 5, the mode spacing increases with stronger cell-to-cell coupling $k$, and will decrease as the number of cells, $N$. As the number of cells goes to infinity, all points on the dispersion curve are filled in.

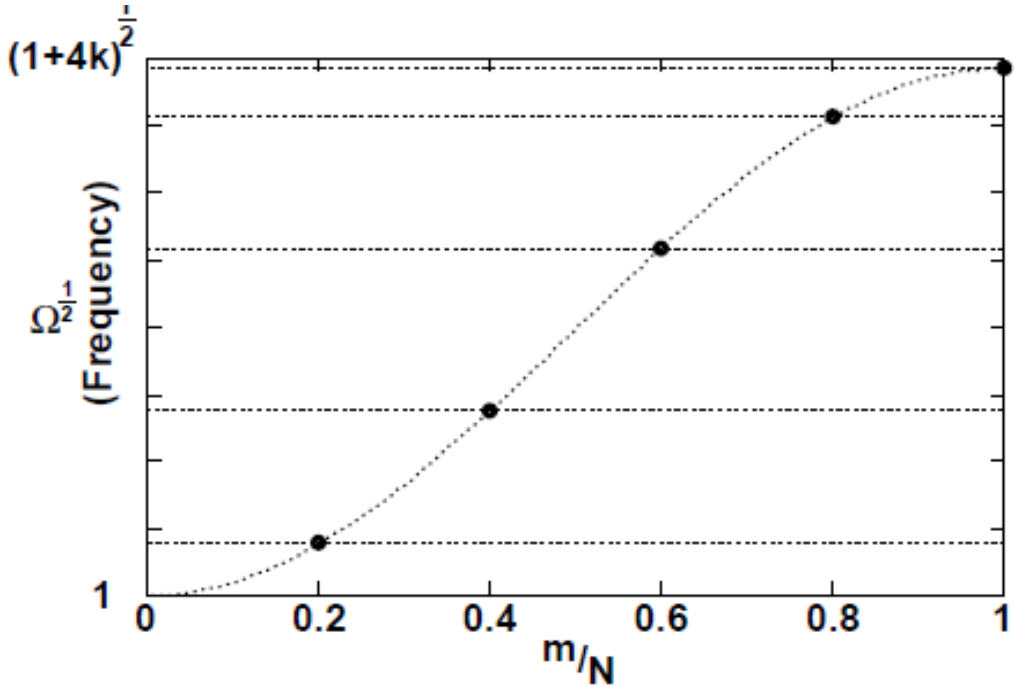

**Fig. 5:** Dispersion relation (frequency vs. mode number) for a five-cell cavity

The cell-to-cell coupling constant *k* can be obtained from the frequencies of the lowest and highest frequency modes via

$$k = \frac{\frac{1}{2}\left[\left(f^{(N)}\right)^2 - \left(f^{(1)}\right)^2\right]}{2\left(f^{(1)}\right)^2 - \left(f^{(N)}\right)^2\left[1 - \cos(\pi/N)\right]}.$$

For a given amount of stored energy in the cavity, it is necessary to have equal fields in each cell so that the net accelerating voltage is maximized and the peak surface EM fields are minimized. This 'flat' field profile is achieved when the cells are properly tuned relative to each other. Cell-to-cell tuning is often needed after initial fabrication when there may be slight deviations in the dimensions of each cell, or after significant etching, or cell deformation due to electron beam welding or heat treatment. The field flatness is measured by perturbing each cell in succession using a small metal (or dielectric) bead while the frequency of the $\pi$ mode is measured. In practice, the bead can be a small (relative to the wavelength) segment of a tube on a fishing line suspended through the cavity along the axis. The relative change in the frequency of each cell is proportional to the relative perturbation of the stored energy and therefore proportional to $E^2$. From the field profile, the tuning parameters can be calculated via the circuit model and perturbation theory [1, 21]. Each cell is then tuned by squeezing or stretching it mechanically.

### 3.3   History of the elliptical cavity shape

Before 1980, multipacting was the dominant limitation in the performance of $\beta = 1$ cavities. Multipacting (MP) stands for '*mult*iple im*pact* electron amplification'. It is a resonant process in which an electron avalanche builds up within a small region of the cavity surface due to a confluence of several circumstances. Electrons in the high magnetic field region travel in quasi-circular orbit segments, returning to the RF surface near to their point of emission, and at about the same phase of the RF period as their emission. Secondary electrons generated upon impact travel along similar orbits. Assuming electrons follow simple cyclotron orbits, a simple rule gives the associated magnetic field for each order of a one-point MP as [22]

$$f/N = eB/2\pi m,$$

where *N* is the order of MP, *e* and *m* are the charge and mass, respectively, of the electron, and *B* is the local magnetic field at the surface. If the secondary emission yield for the electron impact energy is greater than unity, the number of electrons increases exponentially to absorb large amounts of RF power and to deposit it as heat to lower the cavity *Q*. This form of MP is named *one-surface* or *one-point* MP. Depending on the cleanliness of the surface, the secondary emission coefficient of niobium surfaces prepared by cavity treatment methods is larger than unity for electron impact energies between 50 and 1000 eV.

With the invention of the round wall (spherical) cavity shape [20], one-surface MP is no longer a significant problem for velocity-of-light structures. The essential idea is to curve gradually the outer wall of the cavity. Electron trajectories drift toward the equator of the cavity in a few generations (Fig. 6). Near the equator, the electric fields are sufficiently low that energy gain falls well below 50 eV, and regeneration stops because the secondary emission coefficient is less than unity. The same suppression effect is achieved in the elliptical cavity shape, which is generally preferred to the spherical shape due to added mechanical strength and better geometry for rinsing liquids [19].

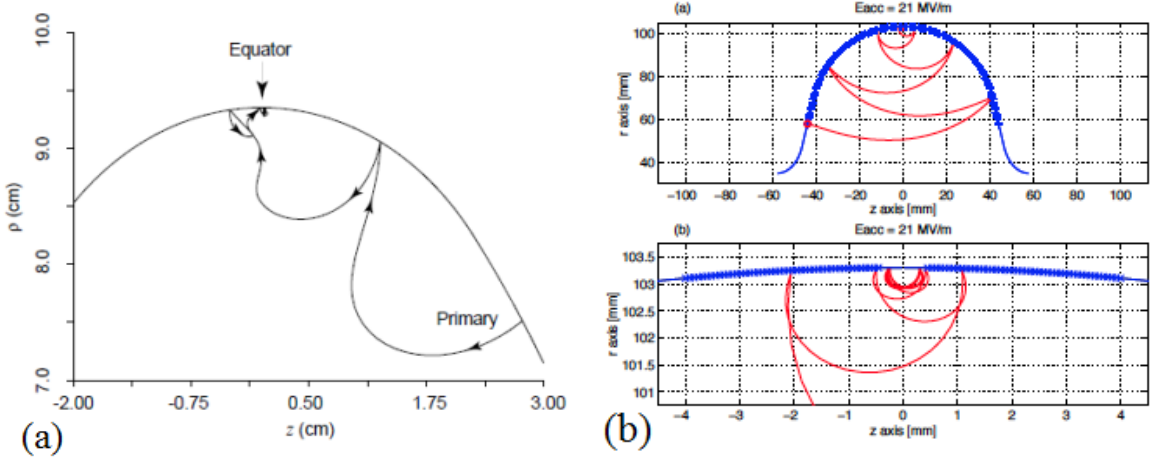

**Fig. 6:** (a) Elimination of one-surface MP by the spherical (elliptical) cell shape. Electrons drift toward the zero electric field region at the equator, where the electric field is so low that the secondary particles cannot gain enough energy to regenerate. (b) Two-point MP in a single-cell 1.3 GHz TESLA-shape cavity near $E_{acc}$ = 21 MV·m$^{-1}$. Note the resonant trajectories in the lower half (expanded).

After one-surface MP was cured, two-point MP was discovered in elliptical cavities when electrons travel to the opposite surface in half an RF period (or in odd-integer multiples of half an RF period). In the spherical/elliptical cavity geometry, two-point MP survives near the equator of the cavity. But the electron energies are low (30–50 eV) near the unity cross-over point of secondary yield, so that the MP is weak and easily processed. The simple rule for two-point MP is

$$2f/(2N-1) = eB/2\pi m.$$

For the elliptical cavities, the peak magnetic field levels of the first-order two-point MP at various frequencies follow the scaling law [22]:

$$B[mT] = 5 + 55 f[GHz].$$

This corresponds to multipacting at 76 mT or $E_{acc}$ = 18 MV·m$^{-1}$ for the TESLA-shape cavity. For general analytic approximations to the fields in the equator region, and the resulting rules for two-point MP in this region, see Ref. [23].

Conditioning times for two-point MP are generally short. During conditioning, MP often grows sufficiently intense to induce a local thermal breakdown of superconductivity. The location of intense MP migrates along the equator as the secondary emission coefficient drops in one place due to electron bombardment. Both MP and its associated breakdown events disappear after sufficient conditioning, but trapped DC magnetic flux generated by thermoelectric currents during the breakdown events reduces the $Q$ values, sometimes by as much as a factor of two. Warming up to >10 K is necessary to remove the trapped magnetic flux. The MP does not reappear since the surface now has a lower secondary yield.

Multipacting levels become suppressed by electron bombardment, which decreases the secondary yield over time, most likely by gas desorption, or possible molecular changes in the adsorbed monolayers. MP can be enhanced again when the secondary emission yield increases due to adsorbates or condensed gas. Levels which have been successfully processed can recur for short periods if the cavity is temperature cycled and gases re-condense on the surface. The occurrence of MP is often accompanied by X-rays when some electrons escape from the MP region into the high electric field region, where they are accelerated.

## 4 Low-velocity structures

### 4.1 Medium-$\beta$

With the growing interest in accelerators for spallation sources, as for example the Spallation Neutron Source [24] SNS at Oak Ridge National Laboratory, elliptical resonators have been extended to high energy (~1 GeV) proton acceleration using medium-$\beta$ superconducting cavities (0.6 < $\beta$ < 0.9). Medium-$\beta$ cavities are also important for high current proton linacs for injectors at Fermilab and CERN, and in the future for energy production via accelerator driven reactors, material irradiation, and nuclear waste transmutation.

The design of a medium-$\beta$ structure involves several tradeoffs. The choice of a low frequency increases the voltage gain per cell, the beam energy acceptance, and the beam quality, at the same time decreasing RF losses and beam losses. But a low RF frequency increases structure size and microphonics level, making RF control more challenging. The larger the number of cells, the higher the voltage gain per structure, but the narrower the velocity acceptance, and the larger the number of cavity designs needed to optimize the voltage gain with changing particle velocity. In the medium velocity range, structures must efficiently accelerate particles whose velocities change along the accelerator. Several structure geometries are therefore needed, each of which is optimized for a particular velocity range. The lower the velocity of the charged particle under acceleration, the faster it will change, and the narrower the velocity range of a particular accelerating structure. This implies that the smaller the value of $\beta$ of a cavity, the smaller the number of cavities of that $\beta$ which can be used in the accelerator.

Efficient acceleration for 0.5 < $\beta$ < 0.9 is achieved in a straightforward manner by axially compressing the dimensions of the standard elliptical resonator geometry while maintaining a constant frequency, as shown in Fig. 7. SNS for example uses two elliptical cavity geometries, one at $\beta$ = 0.6 between 200 MeV and 600 MeV, and the other at β = 0.8 from 600 MeV to 1 GeV [24, 25]. The lower limit of usefulness for the compression approach is about $\beta$ = 0.5, when the vertical flat walls make the structure mechanically unstable.

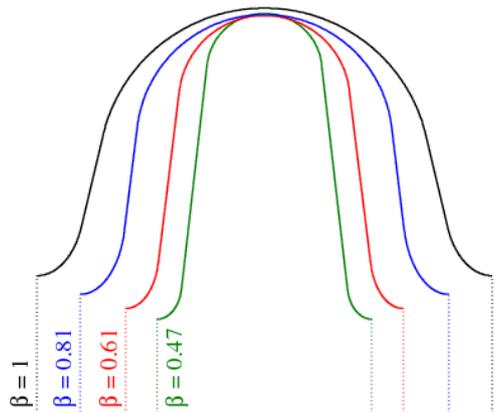

**Fig. 7:** A progression of compressed elliptical cavity shapes at the same RF frequency but for decreasing $\beta$ values [25].

As the cells compress to low $\beta$ geometries, the cavity properties exhibit interesting general trends [26]: $E_{pk}/E_{acc}$ increases from its typical value of 2 to 2.5 for $\beta$ =1 up to 3 or 4 for $\beta$ = 0.5. For 1 MV·m$^{-1}$ accelerating field, the peak surface magnetic field near the equator increases from 4 to 5 mT for $\beta$ = 1 to 6–8 mT for $\beta$ = 0.5. The geometric shunt impedance per cell decreases roughly quadratically as $R/Q$ ($\Omega$) = 120$\beta^2$. For constant $E_{acc}$, the stored energy ($U$) per cell is roughly independent of $\beta$. Structure stored energy plays an important role in amplitude and phase control in the presence of microphonics detuning because the RF power required for phase stabilization depends on the product of the energy content and the amount of detuning. A typical value is $U$ = 200–250 mJ per cell at 1 MV·m$^{-1}$.

## 4.2 Medium-$\beta$ spoke resonators

An alternative path to medium-velocity structures with $\beta$ near 0.5 is via multi-gap spoke resonators (Fig. 8). Here each spoke element is a half-wavelength resonant transmission line operating in a TEM mode. Resonant transmission lines developed originally for low-$\beta$ quarter-wave and half-wave resonator applications will be discussed later.

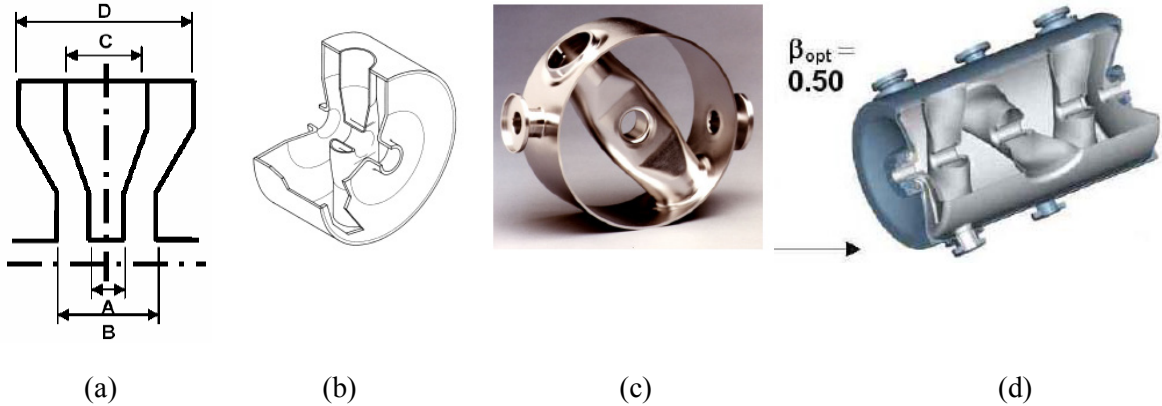

(a)          (b)          (c)          (d)

**Fig. 8:** (a) Spoke and gap profile [27, 28]. (b) 3D sketch [27, 28] and (c) photograph of the first spoke resonator with $\beta$ = 0.28, 800 MHz [29]. (d) Multi-spoke resonator [30].

The spoke elements are made elliptical in cross-section to minimize the peak surface fields. The major axis of the ellipse is normal to the beam axis in the centre of each spoke to minimize the surface electric field and maximize the beam aperture. A typical beam aperture is 4 cm at 345 MHz. In the region of the spokes near the outer cylindrical diameter, the major axis is parallel to the beam axis in order to minimize the peak surface magnetic field. Designs can be optimized by controlling A/B (in Fig. 8(a)) to reduce $E_{pk}/E_{acc}$ and C/D to reduce $B_{pk}/E_{acc}$.

In the spoke structure, the cell-to-cell coupling does not rely on the electric field at the beam holes, as for elliptical cavities, but takes place chiefly via the magnetic field linking cells through the large openings. As a result, the coupling is very strong (20–30% as compared to 2% for $\beta$ = 1 elliptical cavities), which makes the spoke structures robust and the field profiles insensitive to mechanical tolerances. Half end cells (half-gaps) terminate the structure to derive a flat $\pi$ mode.

The range of spoke resonator applications continues to be extended into the medium-$\beta$ regime. In principle there is no clear-cut transition energy from spoke resonators to elliptical ones. Typically TM cavities have an inside diameter of about 0.9$\lambda$. Spoke structures have outer diameters below 0.5$\lambda$. Thus a spoke cavity can be much smaller than an elliptical cavity at the same frequency, or the spoke structure can be made at half the frequency for roughly the same dimensions as the elliptical structure. Choosing a lower frequency allows the option of 4.2 K operation, thus saving capital and operating costs associated with refrigeration.

## 4.3 Low-$\beta$ quarter wave resonators

Low-$\beta$ resonators have been in use for heavy-ion boosters for more than three decades. The short independently phased cavities provide flexibility in operation and beam delivery. Applications continue to expand towards both the lower-$\beta$ as well as the medium-$\beta$ range, as with spoke resonators discussed in Section 4.2. Low-velocity structures must accelerate a variety of ions with different velocity profiles. Different cavity geometries with many gaps have been developed that are suitable for different beam energies, beam currents, and mass/charge ratios.

The Quarter-Wave Resonator (QWR) derives from transmission-line-like elements and belongs to the TEM resonator class. Figure 9 shows a coaxial line, $\lambda/4$ in length shorted at one end to form a resonator with maximum electric field at $\lambda/4$, where the accelerating gaps are located [31]. Low frequencies, typically 100–200 MHz, must be used as the active and useful length of the structure is proportional to $\beta\lambda$. The low frequency results in a large resonator. The typical structure height is about 1 m. The inner conductor, which is made from niobium, is hollow and filled with liquid helium. Operation at 4.2 K is usually possible due to the low RF frequency.

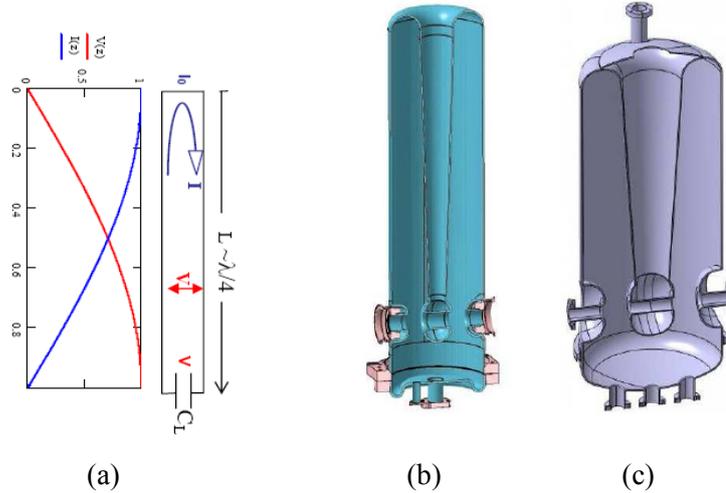

(a) (b) (c)

**Fig. 9:** (a) Equivalent circuit current and voltage distributions of a QWR [31]. (b) Two QWRs for SPIRAL-II with different $\beta$ values (0.07 and 0.12) [32].

The larger the number of gaps in a QWR, the larger the energy gain, but the narrower the velocity acceptance. Figure 10 shows the transit time factor for one-gap and two-gap resonators in the simple approximation of constant field in the gap and zero field outside.

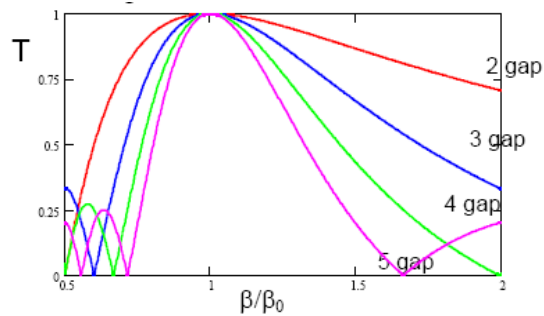

**Fig. 10:** Normalized transit time factor vs. normalized velocity $\beta/\beta_0$, for cavities with different numbers of equal gaps [31].

Being a non-symmetric structure, the QWR has non-symmetrical electromagnetic fields in the beam region; this produces undesirable beam steering through electric and magnetic dipole field components. Compensation can be obtained by gap shaping: the magnetic deflection can be cancelled by enhancement of the electric deflection [33].

Quarter-wave structures are sensitive to mechanical vibrations because of their large size and large load capacitance. The related phase stability is an important issue, particularly for the lowest velocities and for small beam loading, high $Q_{ext}$ operation. The mechanical stability problems have been solved by electronic fast tuners [34] or by the addition of a mechanical damper in the cavity stem [35, 36]. QWRs have covered a wide overall application range: $48 \leq f \leq 160$ MHz, $0.001 \leq \beta \leq 0.2$, with two gaps and four gaps. The extensions to the very-low-$\beta$ regime use a tuning fork arrangement [37]. Compact and modular, QWRs have proven to be efficient high-performance resonators. They can achieve reliably 6 MV·m$^{-1}$.

## 4.4 Half-wave resonators

A half-wavelength ($\lambda/2$) transmission line, with a short at both ends, has maximum voltage in the middle and behaves as a half-wave resonator (HWR). Figure 11 shows a simple equivalent circuit along with voltage and current distributions in one of the $\lambda/2$ loading elements of a single spoke resonator [31].

A HWR is equivalent to two quarter waves facing each other providing the same accelerating voltage as a QWR but with almost twice the power dissipation. Figure 11 shows an example [38]. The symmetry of the structure cancels steering and opens the use of HWRs at $\beta$ from 0.1 to 0.5, above the range customary for QWRs. HWRs also show improved mechanical vibration properties over QWRs.

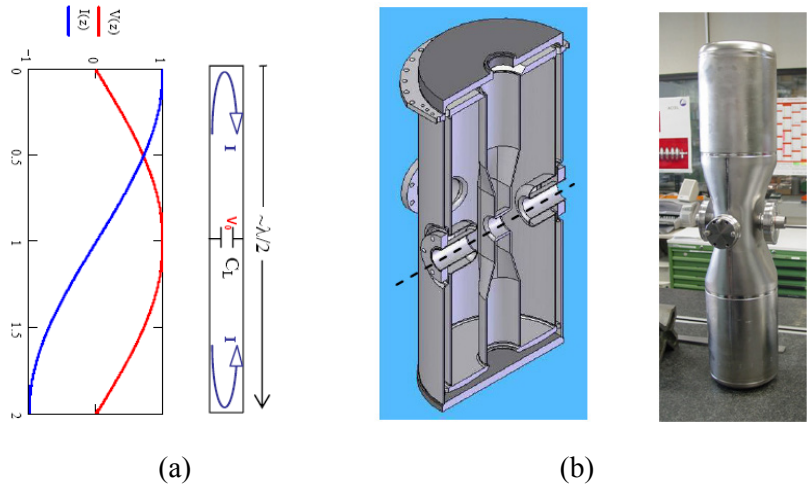

(a) (b)

**Fig. 11:** (a) Equivalent circuit current and voltage distributions of a single-spoke HWR [31]. (b) Example of a HWR [38].

The peak surface electric field occurs at the centre of the loading element. By suitable sizing and shaping of the cross-section, a surface to accelerating field ratio of 3.3 can be obtained, independent of $\beta$. The maximum $H_{pk}$ occurs where the loading element meets the outer enclosure, and is sensitive to the size and shape of the centre conductor. Values of 7 mT·MV$^{-1}$·m$^{-1}$ can be obtained by proper shaping. Most structures are designed with somewhat higher surface field values.

## 5 Mechanical aspects for structure design

The design of a superconducting cavity must take into account several mechanical aspects: stresses, vibrations, and Lorentz forces. Codes exist to simulate mechanical properties [39, 40]. The cavity must withstand stresses induced by the differential pressure between the beam pipe vacuum and atmospheric pressure. Differential thermal contraction due to cool-down from room temperature to cryogenic temperatures induces stress on the cavity walls. Mechanical vibrations of the cavity and the cavity–cryomodule system (microphonics) form another aspect of cavity mechanical design. External vibrations couple to the cavity and excite mechanical resonances, which modulate the RF resonant frequency inducing ponderomotive instabilities [41]. These translate to amplitude and phase modulations of the field, becoming especially significant for a narrow RF bandwidth. Lorentz Force (LF) detuning becomes important in cavity designs for high-field pulsed operations [42]. Surface currents interact with the magnetic field to exert a Lorentz force on the cavity wall. This stress causes a small deformation to change the cavity volume and frequency.

## 5.1 Mechanical stresses

To avoid plastic deformation, the cumulative mechanical stress on the cavity walls must not exceed the cavity material yield strength, including some engineering margin. The frequency shifts due to these stresses must be taken into account when targeting the final frequency or tuner settings and tuner range. Stresses due to the operation of the tuner mechanism should not exceed yield strength while cold. The mechanical requirements may be dealt with by the proper choice of cavity wall thickness or by adding stiffening rings or ribs at locations of high strain.

Medium-$\beta$ elliptical cavities are especially vulnerable to mechanical stresses due to the flattening of the wall, as mentioned.

## 5.2 Vibrations

Stiffeners added at appropriate locations raise the cavity mechanical resonant frequencies so that these no longer couple to the lower-frequency external vibration sources. Dampers introduced in the mechanical system of cavity and cryomodule reduce the mechanical $Q$ of the resonances. The RF bandwidth can also be widened by increasing the strength of the input coupler, but this demands higher RF power and lowers the operating efficiency. Mechanical tuners are usually too slow to counteract cavity wall deformations from microphonics. The stored energy per cell plays an important role in amplitude and phase control in the presence of microphonics detuning. When beam loading is negligible, the amount of RF power required for phase stabilization is given by the product of the energy content and the amount of detuning. Fast tuners of the piezoelectric or magnetostrictive type added to the tuning system provide active damping of microphonics together with sophisticated electronic feedback systems.

## 5.3 Lorentz force detuning

The cavity wall tends to bend inwards at the iris and outwards at the equator, as shown in Fig. 12(a) [43].

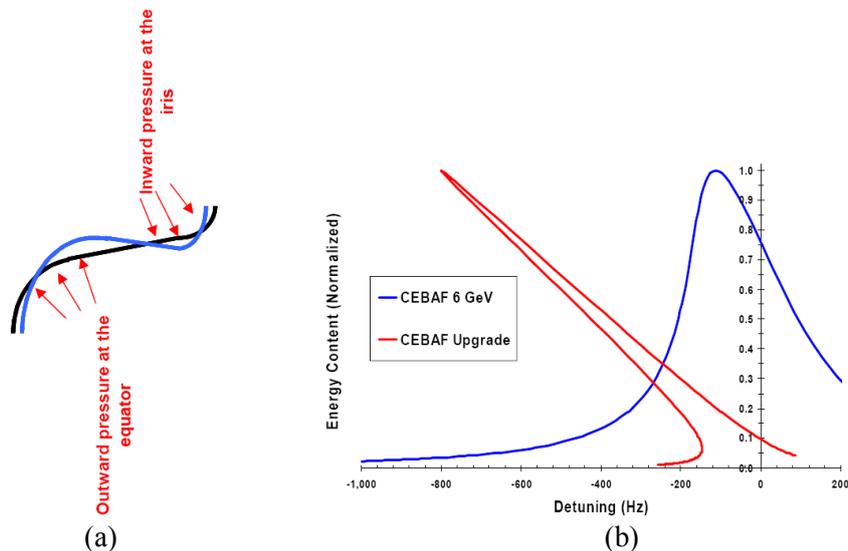

(a)          (b)

**Fig. 12:** Cavity shape distortions due to LF detuning [43, 44]. (a) Lorentz forces acting on different parts of the cavity wall. Note the rotated orientation of the cavity. (b) Distortion of the frequency response of the cavity response at two field levels [45].

The resonant frequency shifts with the square of the field amplitude, distorting the frequency response [45], as shown in Fig. 12(b). Typical detuning coefficients are a few Hz·MV$^{-2}$·m$^{-2}$. This frequency shift can be compensated for by mechanical tuning once the operation field is reached.

A fast tuner is necessary to keep the cavity on resonance, especially for pulsed operation. However, a large LF coefficient can generate 'ponderomotive' oscillations, where small field amplitude errors initially induced by any source (e.g. beam loading) cause cavity detuning through the Lorentz force and start a self-sustained mechanical vibration, which makes cavity operation difficult [46]. LF detuning is especially important in pulsed operation, where the dynamics of the detuning plays a strong role.

Stiffeners must be added to reduce the coefficient, as shown in Fig. 1 [42], but these increase the tuning force. For the TESLA-shape nine-cell elliptical structure (Fig. 1) the LF detuning coefficient is about 2–3 $Hz \cdot MV^{-2} \cdot m^{-2}$, resulting in a frequency shift of several kilohertz at 35 $MV \cdot m^{-1}$, much larger than the cavity bandwidth (300 Hz) chosen for matched beam loading conditions for a linear collider (or XFEL). Stiffening rings in the nine-cell structure reduce the detuning to about 1 $Hz \cdot MV^{-2} \cdot m^{-2}$ [47]. Feedforward techniques can further improve field stability [48-50]. In CW operation at a constant field, the Lorentz force causes a static detuning which is easily compensated for by the tuner feedback, but may nevertheless cause problems during start-up, which must also be dealt with by feedforward in the RF control system.

## 6    Input couplers

### 6.1    Requirements and design principles

An input coupler is a device that efficiently transfers RF power from the generator (source) to a beam loaded cavity by providing a good impedance match between the two, as depicted in the circuit of Fig. 13(a). The coupler must operate over a wide range of load impedance, which varies from a matched impedance at full beam loading to full reflection when there is no beam.

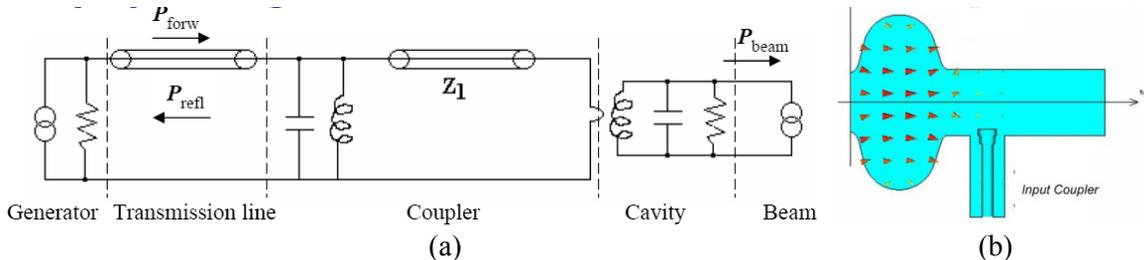

**Fig. 13:** (a) Equivalent circuit for an input coupler. (b) Coaxial input coupler at the beam pipe of a superconducting cavity.

For a superconducting cavity, the input coupler is normally inserted at the beam pipe just outside the end cell of the accelerating structure rather than inside the cell in order to avoid field enhancements that may lower the quench field, or field perturbations that may initiate MP in the cell. The beam pipe diameter and spacing between the end cell and the coupler port need to be sufficient that the antenna does not have to penetrate too far into the beam-line, where it may become a source of strong wakefields or coupler kicks. The transverse electromagnetic field of the power coupler on the beam tube can create a small kick, which increases beam emittance [51]. This effect is especially strong for a cavity at the low-energy end of an accelerator (the injector), where a high average RF power must be coupled to a vulnerable low-energy beam. In this case, a twin coaxial coupler choice reduces harmful transverse kick fields, ideally to zero on-axis. The shape and location of the antenna tip can also be optimized to minimize penetration into the beam pipe.

As an auxiliary device, the coupler design must satisfy numerous requirements and functions. It must preserve the cleanliness of the superconducting cavity, provide a vacuum barrier between the cavity and the feeder waveguide, allow some mechanical flexibility for alignment and thermal contraction during cool-down, permit variable coupling strength (external $Q$) in desired cases, and include thermal transitions from room temperature to cryogenic temperature with minimal static and dynamic thermal losses. In addition, the coupler must be equipped with diagnostic elements to allow safe operation. These requirements call for a careful design from the electromagnetic, mechanical, and thermal points of view.

Couplers for superconducting cavities have been developed to span RF frequencies from 300 to 2000 MHz, and duty factors from 1 to 100%. There has been remarkable progress in power capability for both pulsed and CW operation: 300–500 kW of RF power in operating accelerators and up to 2 MW for prototype testing. There have been many review articles on couplers with valuable references [52–60]. No single coupler design suits all applications. A variety of coupler types have been explored and developed: coaxial and waveguide; one and two windows; cold and warm windows. We will discuss the pros and cons of some of these choices.

There are many reasons for progress in power couplers. Extensive simulations take place in the design phase to predict electromagnetic, thermal, mechanical, and multipacting properties of coupler geometries. Commercial RF modelling codes (e.g., Microwave Studio [61], HFSS [62]) are available for 3D simulations with high accuracy to optimize RF transmission, voltage, current, and power densities. The goal of the RF design is to obtain good transmission properties (minimize reflections and insertion losses) over a workable bandwidth as well as over possible variations in temperature and assembly tolerances. The codes model electromagnetic field distributions over the various elements in the coupler transmission line to establish the best locations for cooling intercepts and window placement, and to determine the coupling strength, normally given in terms of the external $Q$ ($Q_{ext}$). For high-current applications, a good coupler design should also ensure that there are no significant RF fields from higher-order modes that may cause anisotropic heating at the cold window to minimize thermal stresses.

The detailed electromagnetic field distribution is exported into commercial mechanical analysis codes (e.g. ANSYS [39], COSMOS [40]) to calculate stress, vibrations, and heating in regions that bridge ambient and liquid helium temperatures. The goal is to obtain a low cryogenic heat leak by introducing thermal intercepts at proper locations and temperatures. To minimize RF losses, the stainless steel parts of the coupler must be coated with high-conductivity copper of optimal thickness after taking into account the cryogenic heat leak due to conduction, with the goal of minimizing both static and dynamic heat loads overall. In the warm sections of the coupler, the design should avoid a large temperature rise at the operating power level so as to keep manageable the stresses due to thermal expansion and contraction. Cooling designs should take into account the largest possible anticipated thermal load due to operation in travelling and standing wave modes. The mechanical design of the coupler needs to be integrated with the cryomodule design, taking into consideration assembly sequence issues as well as the movement of coupler parts due to cool-down of the module. These shifts (10–20 mm) are usually accommodated by bellows integrated into the mechanical design of the coupler.

Equally important is the implementation of clean practices during fabrication and assembly with high quality control of materials and platings to ensure reliability and high power performance and to preserve cavity cleanliness. Sharp edges should be eliminated in design and fabrication to avoid field enhancement which can lead to field emission. For coated parts, an excellent and reliable bond between film and substrate is essential to stabilize thermally the film, and to prevent particulate generation, which can be dangerous for field emission if such particles fall into the cavity. Use of a cold window is advisable for high gradient applications to seal the cavity from the many coupler components during the early stages of assembly. The cold window should not be in such close proximity to the cavity that impact from field emitted electrons from the cavity lead to window charging, arcing, and possible puncturing. An additional warm window is often used as added protection for vacuum integrity. The space between the two windows must be actively pumped.

Codes are available (see Ref. [63] for a review) to simulate MP in various regions of the coupler to assist in making the best choices for the geometry, for example the inner and outer conductor diameters (and impedance of the coaxial line). In cases where a MP band lies close to an operating point, voltage or magnetic biasing has been developed to disrupt MP resonance conditions for coaxial and waveguide input couplers, respectively. Degassing the coupler by baking keeps it free of surface contamination, thus decreasing the secondary emission and thereby the time required to bring a coupler to the desired power level through the conditioning process.

## 6.2 Power requirements

The input power requirement, $P_f$, is determined by the operating cavity voltage, the beam current, and the RF overhead called for by peak microphonics detuning, and the LF detuning expected [64, 65]:

$$P_f = \frac{V_c^2}{4\frac{R}{Q}Q_{ext}}\left\{\left[1 + \frac{I_b \frac{R}{Q} Q_{ext}}{V_c}\cos\phi_s\right]^2 + \left[\frac{2Q_{ext}\delta\omega_m}{\omega}\right]^2\right\}$$

were $V_c$ is the cavity voltage, $I_b$ is the average beam current, $\phi_s$ is the synchronous phase, $\delta\omega_m$ is the amplitude of the frequency detuning, and $\omega$ is the RF frequency.

Feedback loops provide cavity field stability, reducing the microphonics influence on beam quality as well as the RF power overhead required to compensate for microphonics detuning [64–68]. Environmental microphonic noise creates fluctuations in the cavity resonance frequency and thereby produces amplitude and phase modulations of the field, affecting both beam quality and RF system performance. This is especially true for high-$Q$ superconducting cavities. The optimum $Q_{ext}$ for the power coupler is determined by beam loading:

$$Q_{ext} = V_c^2/[(R/Q)I_b \cos\phi_s].$$

Typical loaded $Q$ for beam-loaded applications range from $10^5$ to several $10^6$. In the case of near zero beam loading, as for example for an Energy Recovery Linac (ERL), the RF power required depends on the microphonics detuning level and choice of $Q_{ext}$ (or loaded $Q$) as shown in Fig. 14 for 1.3 GHz, seven-cell cavity at 20 MV·m$^{-1}$ operating field level.

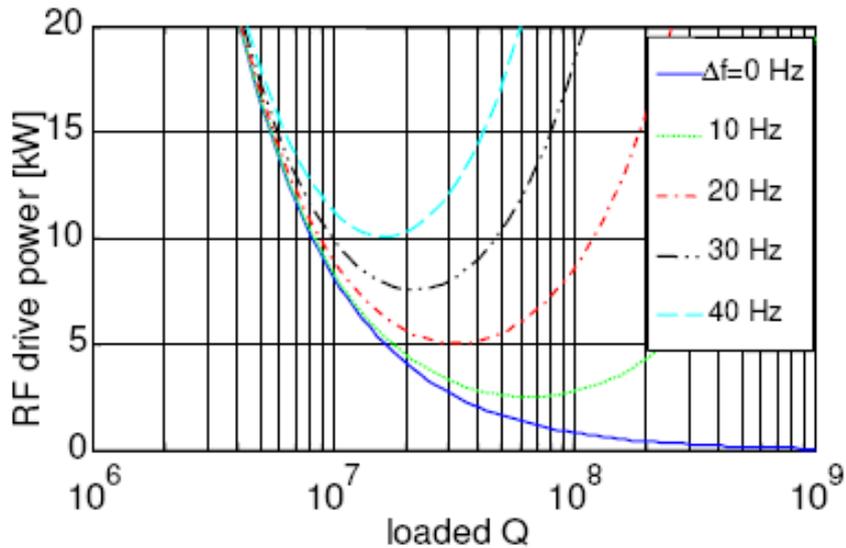

**Fig. 14:** Peak RF drive power as a function of loaded $Q$ ($Q_L$) for a 1.3 GHz, seven-cell cavity at 20 MV·m$^{-1}$ accelerating gradient. The power is determined by the peak microphonics cavity detuning during cavity operation [65].

Although the required power decreases substantially with increasing $Q_L$, running cavities at $Q_{ext}$ in the $10^8$ range is challenging because the small cavity bandwidth of a few hertz makes the RF field extremely sensitive to perturbations of the resonance frequency due to microphonics and LF detuning. Operating at high $Q_L$ makes it hard to meet amplitude and phase stability requirements which can be quite demanding for some applications, such as a high current CW ERL-based light source, where the relative rms amplitude stability must be better than a few × $10^{-4}$ and rms phase <0.1° in order to achieve the beam quality necessary for a good light source. In addition, Lorentz forces during filling detune the cavity by several hundred hertz, making necessary precise compensation during turn-on. Consequently, the highest loaded $Q$s are presently limited to several $10^7$; but RF control advances are forthcoming [65] to lower the RF power requirements for CW accelerators.

Many basic choices need to be made in selecting an RF power coupler design. Among the main factors governing these choices are the RF frequency, the power level, the ease of cooling, the static heat leak, and the coupling adjustability required. Figure 15 compares the two primary varieties: waveguides and coaxial couplers.

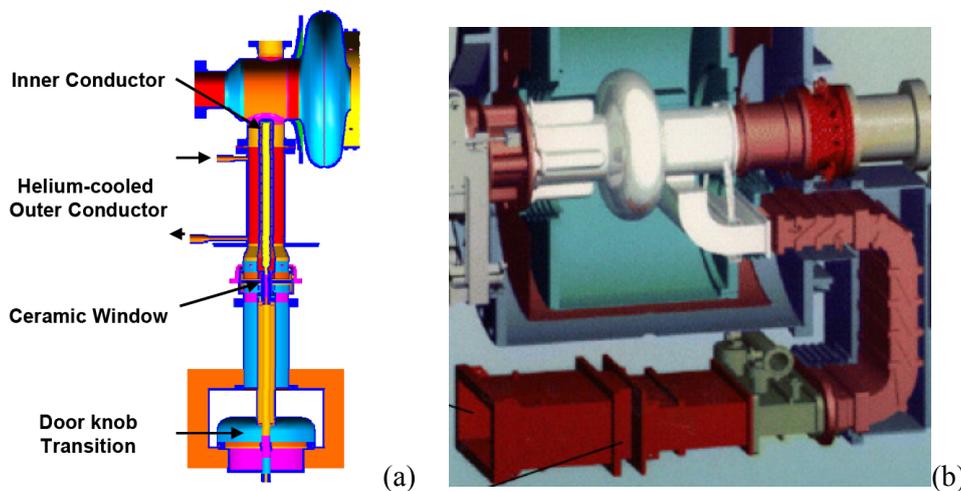

(a)        (b)

**Fig. 15:** (a) Coaxial coupler used for SNS 800 MHz cavities adapted from the KEK-B coupler. Gas flow or water flow through the inner conductor is used for cooling [55]. (b) Waveguide coupler for the CESR 500 MHz cavity used in several storage rings. A planar ceramic disk-shaped window is incorporated in the warm waveguide, which is of reduced height [67, 68].

## 6.3 Waveguide input couplers

Rectangular waveguide couplers are not as widely used as coaxial couplers. Examples of waveguide couplers include CEBAF at 1500 MHz [69] and CESR at 500 MHz [67].

Waveguide coupling is conceptually simpler since it does not require any RF transition between the output waveguide of the RF power source and the cavity interface. Coaxial couplers generally incorporate a transition, such as a door-knob-shaped element. Due to the existence of a cut-off frequency in waveguides, the size of the waveguide coupler is generally larger at a given operating frequency than for the coaxial case. Because of the larger cross-section, there is a larger contribution to the infrared heat transfer to the cryogenic environment. The coupling strength depends on the size and shape of the coupling iris, the location of the waveguide relative to the cavity's end cell, and the location of the terminating short (if any) of the waveguide on the opposite side of the beam tube. Coupling can be adjusted using an external, three-stub tuner waveguide on the air side [70], though the additional field stress and heating due to the standing waves in the line can become problematic for heating and breakdown at high average power levels.

One of the main advantages of the waveguide coupler is the need for cooling only one wall. For 1 MW travelling wave power, the peak electric field for a standard waveguide at 1.3 GHz is

400 kV·m$^{-1}$, whereas for a coaxial line with an outer diameter equal to the small side of the waveguide it is 800 kV·m$^{-1}$ [71]. The power density is lower for the waveguide, but the total longitudinal losses are the same in both cases, about 1 kW·m$^{-1}$ in copper. For the coaxial line, about two-thirds of this loss must be cooled from the inner conductor, which is not as readily accessible as the outer conductor. The losses at the waveguide wall can normally be intercepted at 70 K or 4.5 K, using straps or heat exchanger piping. Waveguides also offer a higher pumping conductance over a coaxial line. MP electrons in the coax can be disrupted by an electrical bias of a few kilovolts [72], whereas MP in the waveguide can be cleared with magnetic bias using a few gauss [73]. However, this approach is not possible in the superconducting waveguide section due to persistent screening currents which exclude dc magnetic flux from the waveguide volume. Here, grooving the waveguide wall is a possible option. The main disadvantage of waveguide couplers is their size, which increases the mechanical and thermal complexity of interfaces to the cavity and cryomodule. Plating and flanging are also harder for rectangular waveguides than for round pipes in coax.

### 6.4 Coaxial input couplers

Not being limited by a cut-off frequency, coaxial couplers are more compact, especially for low-frequency systems. A variety of window geometries and arrangements are available - to be discussed. A large range of coupling values can be achieved by proper insertion of the centre conductor into the line. Also variable coupling can be achieved with a relatively simple adjustment of the inner conductor penetration via a bellows extension. The centre conductor can be electrically isolated from the outer conductor using a kapton film to allow the use of bias voltage to disrupt multipacting. Changing the diameter or impedance of coax lines is a useful method for pushing MP bands to higher power levels [74]. But these have a higher thermal radiation heat leak and a larger interface to the beam tube. The sizing of the coax diameters should avoid azimuthal overmoding.

### 6.5 Windows

A window provides the physical barrier between the cavity vacuum and open waveguide of the power source, but the barrier must be transparent to microwaves at the operating frequency. Many designs use two windows. The main arguments for two windows are (i) to preserve the cleanliness of the cavity by sealing with a first, cold window, and (ii) vacuum safety provided by a second, usually warm window. Superconducting cavities must be handled and maintained under Class 10–100 clean-room conditions at all times to be dust free. It is therefore essential to seal the coupler opening of the cavity with a window at an early stage in the clean room assembly of the input coupler. Placing a window near the cavity allows a compact cavity assembly for ease of handling after sealing in the clean room. Being near the cavity means the window is at 70 K or lower, and therefore must have a vacuum on both sides. Hence the window can be cooled only by conduction, making high average power design more challenging. Multipacting can occur in the vacuum on both sides of the cold window. If the cold window is too close to the cavity field, emission electrons from the cavity can charge it up, leading to arcing and eventually ceramic damage [75, 76].

The second window prevents gas condensation on the cold window, and serves as a backup to preserve the cavity vacuum in case the cold window develops a leak during operation. The vacuum between the windows must be pumped separately. The second window is normally incorporated into the transition from coaxial to waveguide. It can also be a planar waveguide window or coaxial disk window. The warm part of the coupler, including the second window, is generally assembled after placing the cavity string into the vacuum vessel, also under clean and dry conditions for faster processing to high power. Cooling designs for both windows should take into account the largest possible anticipated thermal load due to operating the window and coupler in a full standing wave condition swept through 180° phase change.

The cold window design is a must for applications aiming for the highest gradients (>20 MV·m$^{-1}$), to prevent dust contamination and field emission during subsequent assembly steps. For high (≥100 kW) average power applications at moderate to low gradients (5–20 MV·m$^{-1}$), a single, warm window design is often used with convection cooling or water cooling. A gas barrier serves as the second window to provide safety for the cavity vacuum in case the main window develops a leak. In this case, the cavity is exposed only to the dry, dust-free air in between the two windows. The warm window is located sufficiently far from the cavity cold mass to limit both conductive and radiation heat leaks into the liquid helium bath. The challenge for the single warm window design is that a large coupler assembly must be attached to the clean cavity in a clean room. Several types of ceramic windows are in use. Coaxial couplers use either the cylindrical window [77] or disk window [78]. Waveguide couplers generally use a planar rectangular ceramic [79] incorporated within the rectangular waveguide.

## 6.6 Prime example of a coaxial coupler

Figure 16 shows the geometry of TTF-III (third generation), 1.3 GHz coupler developed at the Tesla Test Facility (TTF) used for FLASH and the XFEL and possibly the ILC [80]. The coupler is designed for operation in the pulsed mode (1.3 ms) at several hundred kilowatt power and less than 5 kW average power. It has an adjustable coupling strength between $1 \times 10^6$ and $2 \times 10^7$ for 15 mm antenna movement. There are two cylindrical windows (97.5% $Al_2O_3$ with TiN coating), one at 70 K and one warm near the door-knob transition. The cold part seals the cavity vacuum, and is entirely inserted into the cryomodule. The warm part has its own separate vacuum. The cold coaxial line has 70 Ω impedance with 40 mm Outer Diameter (OD), and the warm line has 50 Ω impedance with 62 mm OD. All stainless steel parts are made of 1.44 mm thick tubes. Copper plating is 30 $\mu$m thick on the inner conductor and 10 $\mu$m thick on the outer conductor. There are two heat intercepts: at 4.2 K and at 70 K.

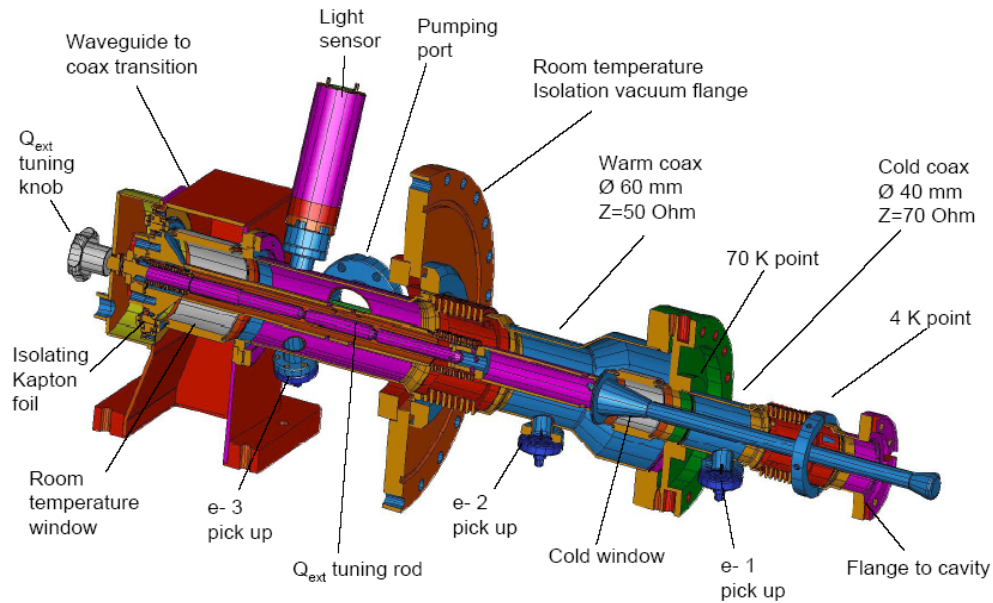

**Fig 16:** TheTTF-III coupler

A prime example of the waveguide version is the CESR SRF waveguide coupler (Fig. 15(b)) [67]. It has a fixed coupling, $Q_{ext} = 2 \times 10^5$, with a factor of three adjustability via a three-stub waveguide transformer. Magnetic bias by solenoids wound around the normal conducting waveguide sections helps to suppress MP.

# 7 Higher-order modes and couplers

## 7.1 Higher-order modes

When passing through an accelerating cavity, a particle beam excites a wide spectrum of higher-modes, depending on the impedances (i.e. the *R*/*Q*) of those modes. The resulting electromagnetic field left behind by the beam is called the wakefield. Thus the passage of the beam can deposit significant power in high impedance, monopole Higher-Order Modes (HOMs). Unless properly extracted and damped, HOMs can also cause longitudinal beam instabilities and increase the beam energy spread. The energy lost by the passage a single bunch, charge *q*, is given by

$$U_q = k_n q^2, \quad k_n = \frac{\omega_n}{4} \frac{R_{sh}}{Q_0},$$

where $\omega_n$ is the angular frequency of mode *n*, $R_{sh}/Q_0$ is the geometric shunt impedance of the monopole; $k_n$ is also referred to as the loss factor of mode *n*. The total power deposited depends on the number of bunches per second, or the beam current.

Among the deflecting modes, dipoles have the highest impedance. The energy lost by a charge to the dipole mode is given by

$$U_q = k_d q^2 \left(\frac{\rho}{a}\right)^2, \quad k_d = a^2 \left(\frac{\omega_n}{c}\right)^2 \frac{\omega_n}{4} \frac{R_d}{Q_0},$$

where $\rho$ is the bunch displacement off-axis, *a* is the cavity aperture (radius), $\omega_n$ is the angular HOM frequency of mode *n*, and $R_d/Q_0$ is the dipole mode impedance, formally defined in Ref. [1]. Each dipole mode has two polarizations split with a small frequency difference due to perturbations, such as the presence of couplers. Dipole modes with high transverse *R*/*Q* are harmful for emittance growth of the beam.

## 7.2 HOM couplers

The main functions of HOM couplers are to remove the beam-induced power in the monopole HOMs and to damp the dangerous monopole and dipole modes to avoid energy spread, beam emittance degradation, and beam blow-up after multiple beam passages. The beam-induced HOM power in monopole modes must be extracted from the cavity and deposited to higher temperatures to avoid cryogenic losses. Modes with high shunt impedance and high *Q* (i.e. high *R/Q\*Q*) are of particular concern.

As a prime example for monopole mode excitation and power deposition, consider the European XFEL, which plans to use the TESLA-shaped cavity developed by TTF. The dangerous modes are damped by two antenna/loop couplers (Fig. 17) placed just outside the end cells on either end of the cavity [15]. In general, the coaxial HOM coupler has an antenna or pick-up loop to extract HOM energy from the end cell via the electric or magnetic fields. Inductive and capacitive elements within the coupler body enhance coupling at the desired frequencies of high *R/Q* modes, and suppress coupling to the fundamental mode via a rejection filter with a large rejection ratio, typically more than –70 dB. The rejection filter must be carefully tuned prior to cavity installation, and the transmission line must be terminated by a broad-band load. For high average current application, the couplers are located inside the helium vessel for good cooling.

The measured damping of these modes is shown in Fig. 18(a) [81]. Antenna/loop couplers on the beam pipe optimized to damp the low-frequency high-impedance modes are generally not as effective against a broad spectrum of propagating HOMs with frequencies above the cut-off frequency of the beam pipe. Hence a broad-band beam pipe absorber is needed for applications with a short bunch. The damping for dipole modes is shown in Fig. 18(b) [81]. Most of the dangerous dipole modes are well damped relative to the beam dynamics requirement of $Q_{ext} \sim 10^5$.

As with input couplers, HOM couplers must also be placed outside the cells to avoid field enhancement in the cells, which may lead to premature quench or multipacting. However, some HOMs have very little stored energy in the end cells due to mode 'trapping' within the central cells of a many-cell structure. Their suppression becomes difficult. Reducing the number of cells and/or enlarging the iris diameter minimizes the likelihood of trapping by enhancing the cell-to-cell coupling. Another approach to minimize trapping is to match the end- and inner-cell frequencies by adjusting the shape of the end cells, which yields a slightly asymmetric cavity.

Antenna/loop-based HOM couplers also introduce kicks, which can spoil the beam emittance especially at low beam energy. These kicks arise from the wakefields introduced by the coupler geometry as well as by RF fields and the asymmetry of the coupler locations with respect to the beam axis. The kicks are reduced by symmetrizing the placement of multiple HOM couplers.

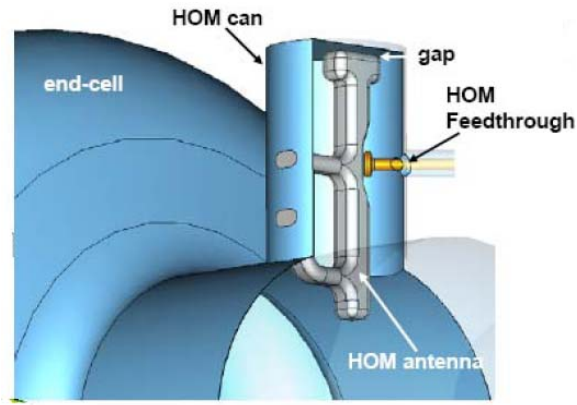

**Fig. 17:** The TESLA HOM coupler

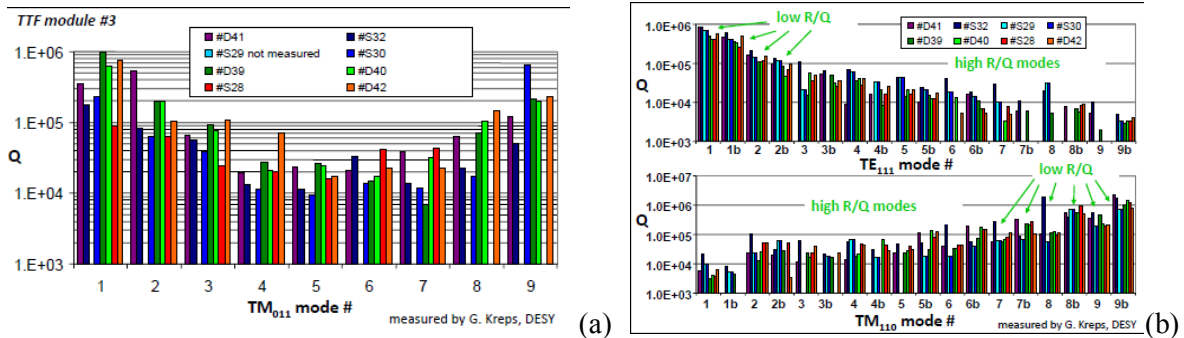

**Fig. 18:** (a) Damping of dangerous monopole HOMs. (b) Damping of the dangerous dipole HOMs

The TESLA HOM coupler was scaled to 805 MHz for SNS at 6% duty cycle operation, and to 1500 MHz for the CEBAF 12 GeV upgrade and CW operation, and also to 3.9 GHz for the third-harmonic TTF injector cavity [82, 83]. These higher average power applications have met with some difficulties so that modifications had to be developed. The problem arises due to the heating of the output antenna by the residual magnetic field of the fundamental mode (several percent of the field on equator) and the heat leak of the output line. Abnormal heating of HOM couplers can detune the notch filter and couple out substantial amounts of power from the fundamental mode, leading to thermal runaway. Other causes for abnormal heating observed are multipacting in the HOM coupler, and heating from the impact of field emitted electrons emanating from the cavity, or from neighbouring cavities. Enabling higher duty factor operation (or higher damping) by bringing the coupler tip closer to the end cell requires improvements in cooling. There is a significant amount of stored energy in the transmission line coupler. The high electric field regions of the loop coupler are also susceptible to multipacting and associated heating. The troublesome regions are between the loop and the wall, in the small gap which defines the notch filter, between the coaxial post and the end wall of the can, and at several places between the post and the cylindrical wall.

Microwave analysis combined with thermal analysis using codes such as HFSS and ANSYS have been used to analyze heating difficulties and devise solutions [84, 85]. To keep the output antenna superconducting, one approach has been to shorten the antenna probe tip, provided the HOM coupling loss can be tolerated. Another is to enhance the heat conduction at the output connector, for example by using a larger RF feed-through with sapphire window and cooling copper blocks [86]. Shortening the antenna tip is one way to reduce the fields and suppress MP at the required field levels.

## 7.3 Waveguide HOM couplers

Waveguides as HOM couplers provide convenient and natural high pass filters that reject the fundamental mode and require no fine tuning, unlike the fundamental mode rejection filter essential to coaxial couplers. The waveguide coupler removes HOMs over a wide frequency range. It can handle high HOM power without heating difficulties. However, the bulkier waveguide adds to structure cost of the HOM-end groups.

The first waveguide HOM couplers for superconducting cavities were developed at Cornell for 1.5 GHz muffin-tin cavities, and subsequently adopted for the 1.5 GHz Cornell/CEBAF elliptical five-cell cavities (Fig. 19). The several milliamp beam current of CEBAF results in a small amount of the HOM power to allow termination of the HOM couplers with waveguide loads inside cryomodules [87]. In case of higher beam current, the terminating loads must be located outside the cryomodule, which adds to the mechanical complexity of the waveguide option and introduces additional cryogenic heat leaks and shielding requirements.

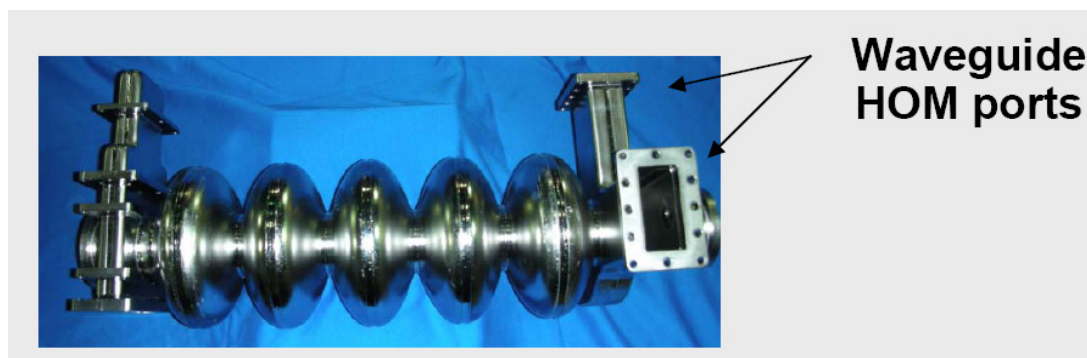

**Fig. 19:** Waveguide HOM coupler examples for the Cornell/CEBAF cavity

## 7.4 Beam pipe couplers and absorbers

The cavity beam pipe can be viewed as a transmission line to couple out HOMs with frequencies above the beam pipe cut-off. The fundamental mode is below cut-off and does not propagate out, providing a natural rejection filter. As with the waveguide coupler, no tuning is needed. The cylindrical symmetry of the beam pipe avoids coupler kicks. The diameter of the beam pipe can be chosen to couple out all monopole and dipole modes. Extraction of the first two dipole modes demands the largest opening. These high-impedance modes are especially dangerous. But a large diameter beam pipe reduces the rejection and $R/Q$ of the fundamental mode and enhances the peak surface fields for operation. These effects can be reduced by introducing a rounded step in the beam pipe at the iris of the end cell. In some cases, as for KEK-B [88], the largest beam pipe is installed at only one end of the cavity for the extraction of the first two dipoles. A section of the beam pipe lined with a microwave absorbing material serves as the load. This can be placed at room temperature outside the cryostat, or at ~80 K inside. But the presence of numerous absorbing sections along the beam line reduces the overall real-estate gradient.

Beam-line HOM couplers (Fig. 20) are especially suitable for high-current, short bunch accelerators. The development of these are reviewed in Ref. [1]. Several storage rings now use the approach: CESR at Cornell, KEK-B in Tsukuba, Taiwan Light Source, Canadian Light Source, DIAMOND light source, and BEPC-II in Beijing. The HOMs are damped to $Q$ values between 100 and 1000. Measurements of the electromagnetic properties of absorbing materials have been performed.

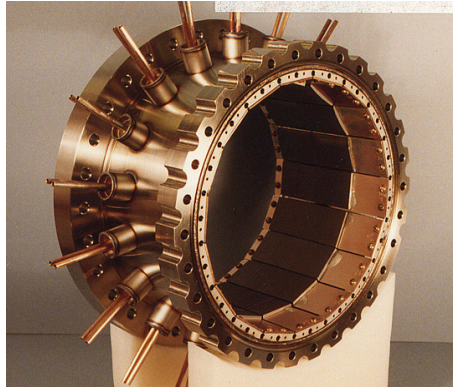

**Fig. 20:** Beam-pipe absorbers lined with ferrite [89]. The CESR load with 3 mm TT211R ferrite tiles bonded to a sintered copper–tungsten plate with Ag–Sn alloy.

## 8    Tuners

Frequency tuners are an essential component of acceleration systems. Both slow tuners and fast tuners fulfil important functions. Recent review talks can be found in Refs. [90–92]. Slow tuners bring a cavity resonance to the operating frequency, compensating for a variety of effects: cavity dimensional changes due to evacuation and cool-down, or slow drifts in frequency due to pressure changes in the helium bath surrounding the cavity. Tuners also compensate for the reactive effects of beam loading in high-current accelerators to minimize the reflected power. Occasionally cavities need to be detuned to bypass operation, or for diagnostic purposes.

Slow tuners must cover a wide tuning range (of up to several hundred kilohertz), while providing a resolution of the order of 1 Hz. Slow tuners are usually motor driven. Fast tuners provide a smaller tuning range of several cavity bandwidths, but with a control bandwidth of several kilohertz and slew rates of 1 $\mu$m in 100 $\mu$s. Together with feed-forward and feedback, these tuners compensate for static and dynamic LF detuning, especially at high-gradient operation [93]. Fast tuners also have the potential to control microphonics, typically up to several tens of hertz. Fast tuners are important for cavities with little beam loading, when operation at high $Q_{ext}$ is desirable to minimize RF power. At high $Q_{ext}$, the bandwidth is sufficiently narrow that microphonic excitations disrupt the cavity resonance. Typical microphonic noise levels are of the order of few hertz to several tens of hertz with a frequency spectrum ranging up to a few hundred hertz. The observed spectrum is a result of a convolution of the spectrum of excitation and the coupling to the mechanical resonances of the cavities. Typical excitation sources of microphonics are vibrations from pumps and human activity. If the repetition frequency of the coarse tuner stepping motor matches a mechanical resonance of the cavity–cryostat system, strong mechanical vibrations can be excited. To avoid microphonic excitations in general, it is important to ensure that the mechanical resonant frequencies of the structure do not coincide with the frequency of RF repetition rate

Fast tuners are generally integrated with the slow mechanical tuner and mostly based on piezoelectric elements [94, 95]. The typical static tuning range is 1 kHz or less.

Tuner designs strive for compactness to avoid wasting beam-line space, disrupting the field flatness from cell–cell, or tuning neighbouring cavities. The tuner mechanical supports and operating motors should keep cryogenic heat load to a minimum. Frequency tuners should be free of hysteresis. Pre-setting is required to avoid the neutral point between tension and compression over the entire expected range of operation. The hardware must be easy to maintain and repair, ideally without the need to warm up or disassemble a module, but in practice this has been realized only for a few tuner choices.

A variety of tuner designs are now available as a result of inventive efforts at a number of laboratories. One such tuner used at FLASH and the XFEL, called the Saclay/TTF tuner, is shown in Fig. 21.

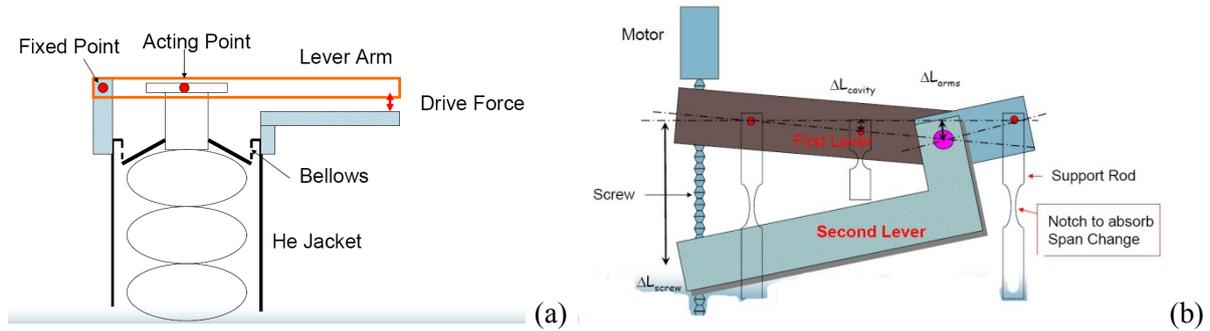

**Fig. 21**: (a) Basic principle of most slow tuners [96]. (b) Principle of the Saclay tuner used in TTF [97]

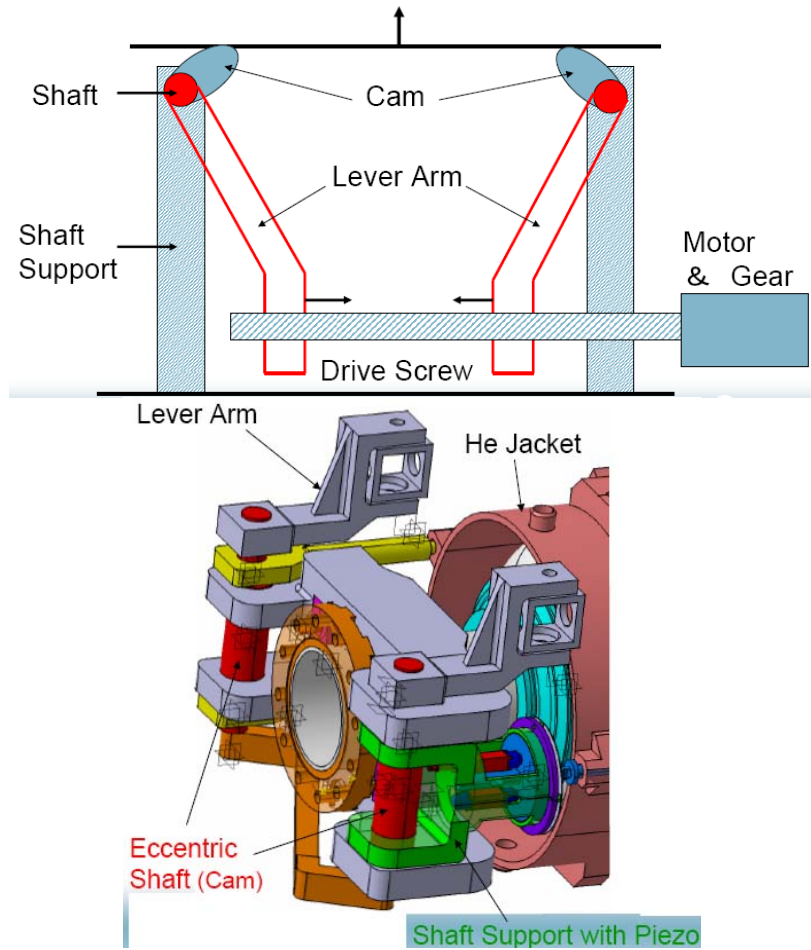

**Fig. 22** (a) Principle of lever–cam design of the improved Saclay tuner [96]. (b) 3D CAD layout [97]

**References**

[1] H. Padamsee, J. Knobloch and T. Hays, *RF Superconductivity for Accelerators* (Wiley & Sons, New York, 1998).

[2] H. Padamsee, *RF Superconductivity: Science, Technology and Applications* (Wiley-VCH, Weinheim, 2009).

[3] P. Schmueser, Basic principles of RF superconductivity and superconducting cavities, Proc. 11th Workshop on RF Superconductivity, Travemünde, Germany, 2003, paper MoTo 1.

[4] H. Padamsee, in *Frontiers of Accelerator Technology*, Eds. S.I. Kurokawa *et al*. (World Scientific, Singapore, 1996), p. 383.

[5] D. Proch, in *Handbook of Accelerator Physics and Engineering* (World Scientific, Singapore, 1999), p. 530.

[6] S. Belomestnykh, *Rev. Accel. Sci. Technol.* **5** (2012) 147.

[7] M. Kelly, *Rev. Accel. Sci. Technol.* **5** (2012) 185.

[8] M.P. Kelly, in *Handbook of Accelerator Physics and Engineering* (World Scientific, Singapore, 2013), p. 681.

[9] H. Padamsee, in *Encyclopedia of Electrical and Electronics Engineering*, Ed. J.G. Webster (John Wiley and Sons, New York, 1999).

[10] D. Proch, *Rep. Prog. Phys*. **61** (1998) 431.

[11] H. Padamsee, Accelerating applications of RF superconductivity—success stories, Proc. 2004 Applied Superconductivity Conf., Jacksonville, FL, 2004, p. 2432.

[12] W. Weingarten, Superconducting cavities—basics, Proc. Joint US-CERN-Japan Int. School, Frontiers of Accelerator Technology, Eds. S.I. Kurokawa *et al*. (World Scientific, Singapore, 1994), p. 311.

[13] H. Padamsee and J. Knobloch, Issues in superconducting RF technology, Proc. Joint US-CERN-Japan Int. School, Frontiers of Accelerator Technology, Eds. S.I. Kurokawa *et al*. (World Scientific, Singapore, 1994), p. 101.

[14] E. Haebel *et al*., Cavity shape optimization for superconducting linear collider, Proc. HEACC 1992, High Energy Accelerator Conference, Hamburg, 1992, p.957.

[15] B. Aune *et al*., *Phys. Rev. ST Accel. Beams* **3** (2000) 092001.

[16] H. Padamsee *et al*., *Part. Accel*. **40** (1992) 17.

[17] H. Padamsee *et al*., Accelerating cavity development for the Cornell B-Factory, CESR-B, Proc. PAC1991, Particle Accelerator Conf., San Francisco, CA, 1991, vol. 2, p. 786.

[18] T. Furuya *et al*., Proc. 7th Workshop on RF Superconductivity, Gif-sur-Yvette, France, 1995, p. 729.

[19] P. Kneisel *et al*., *Nucl. Instrum. Meth. Phys. Res*. **188** (1981) 669.

[20] U. Klein and D. Proch, Multipacting in superconducting RF structures, Proc. Conf. Future Possibilities for Electron Accelerators, Charlottesville, NC, 1979, p. N1.

[21] P. Schmueser, Tuning of Multi-cell cavities using bead-pull measurements, SRF920925-10, Cornell University Internal Report.

[22] R.L. Geng, Multipacting simulations for superconducting cavities and RF coupler waveguides Proc. PAC2003, Particle Accelerator Conf., Portland, Oregon, 2003, p. 264.

[23] V. Shemelin, *Phys. Rev. ST Accel. Beams* **16** (2013) 012002.

[24] S.-H. Kim, SNS superconducting linac operational experience and upgrade path Proc. LINAC'08 Linear Accelerator Conference, Victoria, British Columbia, Canada, 2008, p. 11.

[25] C.C. Compton *et al*. *Phys. Rev. ST Accel. Beams* **8**(4) (2005).

[26] J.R. Delayen, Medium-$\beta$ superconducting accelerating structures, Proc. 10th Workshop on RF Superconductivity, Tsukuba, Japan, 2001.

[27] J.R. Delayen, Medium-$\beta$ superconducting accelerating structures, USPAS, Baton Rouge, LA, 2003.

[28] J.R. Delayen, Low and medium $\beta$ cavities and accelerators, Proc. 13th Workshop on RF Superconductivity, Beijing, China (2007), tutorial 4a.

[29] K.W. Shepard *et al.*, Prototype 350 MHz, niobium spoke-loaded cavities, Proc. PAC 1999, Particle Accelerator Conference, New York, NY, 1999, p. 955.

[30] K.W. Shepard *et al.*, Development of spoke cavities for RIA, Proc. 12th Workshop on RF Superconductivity, Ithaca, NY, 2005, p. 334.

[31] A. Facco, Tutorial on low beta cavity design, Proc. 12th Workshop on RF Superconductivity, Ithaca, NY, 2005, p. 21.

[32] G. Devanz, SPIRAL 2 resonators, Proc. 12th Workshop on RF Superconductivity, Ithaca, NY, 2005, p. 108.

[33] P.N. Ostroumov and K.W. Shepard, *Phys. Rev. ST. Accel. Beams* **11** (2001) 030101.

[34] N. Added *et al.*, Upgraded phase control system for superconducting low velocity accelerating structure, Proc. LINAC 1992, Linear Accelerator Conference, Ottawa, Ontario, 1992, p. 181.

[35] A. Facco *et al.*, On-line performance of the LNL mechanically damped superconducting low beta resonators, Proc. EPAC 1998, European Particle Accelerator Conference, Stockholm, 1998, p. 1846.

[36] A. Facco, *Part. Accel.* **61** (1998) 265.

[37] K.W. Shepard, Superconducting low-velocity linac for the Argonne positive-ion injector, Proc. PAC 1989, Particle Accelerator Conference, Chicago, IL, 1989, p. 1974.

[38] M. Pekeler, et al, Performance of a Prototype 176 MHz $\beta$= 0.09 Half-Wave Resonator for the SARAF Linac, Proc. 12[th] Workshop on RF Superconductivity, Ithaca, NY, USA, p. 331 (2005).

[39] ANSYS, Inc., Southpointe, Canonsburg, PA, http://www.ansys.com

[40] COSMOS, Structural Research and Analysis Corp., Santa Monica, CA, http://www.cosmosm.com

[41] J.R. Delayen, Ponderomotive instabilities and microphonics, a tutorial, *Phys. C* **441** p. 1 (2006).

[42] J. Sekutowicz, Superconducting high-$\beta$ cavities, Proc. 13th Workshop on RF Superconductivity, Beijing, China, 2007, tutorial 2a.

[43] R. Mitchell *et al.*, Lorentz force detuning analysis of the Spallation Neutron Source (SNS) accelerating cavities, Proc. 10th Workshop on RF Superconductivity, Tsukuba, Japan, 2001, p. 236.

[44] J.R. Delayen, LLRF control and tuning systems, Proc. 13th Workshop on RF Superconductivity, Beijing, China, 2007, tutorial 4b.

[45] J.R. Delayen, Low and medium $\beta$ cavities and accelerators, Proc. 13th Workshop on RF Superconductivity, Beijing, China, 2007, tutorial 4a.

[46] P. Brown *et al.*, Operating experience with the LEP2 superconducting RF system, Proc. 10th Workshop on RF Superconductivity, Tsukuba Japan, 2001, p. 185.

[47] M. Liepe *et al.*, Dynamic Lorentz force compensation with a fast piezoelectric tuner, Proc. PAC 2001, Particle Accelerator Conference, Chicago, IL, 2001, p. 1074.

[48] S. Simrock, Advances in RF control for high gradients, Proc. 9th Workshop on RF Superconductivity, Santa Fe, NM, 1999, p. 92.

[49] S. Simrock, Achieving phase and amplitude stability in pulsed superconducting cavities, Proc. 10th Workshop on RF Superconductivity, Tsukuba, Japan, 2001, p. 231.

[50] A. Brandt *et al.*, General automation of LLRF control for superconducting accelerators, Proc. 12th Workshop on RF Superconductivity, Ithaca, NY, 2005 [*Phys. C* **441** (2006) 263].

[51] V. Shemelin *et al.*, Dipole-mode-free and kick-free 2-cell cavity for the SC ERL injector, Proc. PAC 2003, Particle Accelerator Conference, Portland, OR, 2003, p. 2059.

[52] M.S. Champion, RF input couplers and windows: performances, limitations, and recent developments, Proc. 7th Workshop on RF Superconductivity, Gif-sur-Yvette, France, 1995, p. 195.

[53] D. Proch, Techniques in high-power components for SRF cavities, a look to the future, Proc. LINAC2002, Linear Accelerator Conference, Gyeongju, Korea, 2002, p. 529.

[54] B. Rusnak, RF power and HOM coupler tutorial, Proc. 11th Workshop on RF Superconductivity, Travemünde, Germany, 2003, p. 496.

[55] I. Campisi, Fundamental power couplers for superconducting cavities, Proc. 10th Workshop on RF Superconductivity, Tsukuba, Japan, 2001, p. 132.

[56] I. Campisi, State of the art power couplers for superconducting RF cavities, Proc. EPAC 2002, European Particle Accelerator Conference, Paris, France, 2002, p. 144.

[57] S. Belomestnykh, Review of high power CW couplers for superconducting cavities, Proc. Workshop on High-Power Couplers for Superconducting Accelerators, Newport News, VA, 2002, http://www.jlab.org/intralab/calendar/archive02/HPC/papers.htm

[58] S. Belomestnykh, Overview of input power coupler developments, pulsed and CW, Proc. 13th Workshop on RF Superconductivity, Beijing, China, 2007, paper WE305.

[59] T. Garvey, The design and performance of CW and pulsed power couplers – a review, Proc. 12th Workshop on RF Superconductivity, Ithaca, NY, 2005 [*Phys. C* **441** (2006) 209].

[60] A. Variola, High power couplers for linear accelerators, Proc. LINAC 2006, Linear Accelerator Conference, Knoxville, TN, 2006, p. 531.

[61] CST Microwave Studio, CST GMbH, Darmstadt, Germany, http://www.cst.com/Content/Products/MWS/Overview.aspx

[62] HFSS, Ansoft Corp., Pittsburgh, PA, http://www.ansoft.com/products/hf/hfss/

[63] F. Krawczyk, Status of multipacting simulation capabilities for SCRF applications, Proc. 10th Workshop on RF Superconductivity, Tsukuba, Japan, 2001, p. 108.

[64] M. Liepe, Microphonics detuning in the 500 MHz superconducting CESR cavities, Proc. PAC 2003, Particle Accelerator Conference, Portland, OR, 2003, p. 1326.

[65] M. Liepe *et al.*, Pushing the limits: RF field control at high loaded Q, Proc. PAC 2005, Particle Accelerator Conference, Knoxville, TN, 2005, p. 2642.

[66] M. Liepe, RF parameter and field stability requirements for the Cornell ERL prototype, Proc. PAC 2003, Particle Accelerator Conference, Portland, OR, 2003, p.1329.

[67] S. Belomestnykh and H. Padamsee, Performance of the CESR superconducting RF system and future plans, Proc. 10th Workshop on RF Superconductivity, Tsukuba, Japan, 2001, p. 197.

[68] S. Belomestnykh *et al.*, Superconducting RF system upgrade for short bunch operation of CESR, Proc. PAC 2001, Particle Accelerator Conference, Chicago, IL, 2001, p.1062.

[69] J.R. Delayen *et al.*, An RF input coupler system for the CEBAF energy upgrade cryomodule, Proc. PAC 1999, Particle Accelerator Conference, New York, NY, 1999, p. 1462.

[70] V. Veshcherevich and S. Belomestnykh, Correction of the coupling of CESR RF cavities to klystrons using three-post waveguide transformers, Report SRF020220-02, Laboratory for Elementary-Particle Physics, Cornell University (2002).

[71] D. Proch, Techniques in high-power components for SRF cavities, a look to the future, Proc. LINAC2002, Linear Accelerator Conference, Gyeongju, Korea, 2002, p. 529.

[72] J. Tuckmantel *et al.*, Improvements to power couplers for the LEP2 superconducting cavities, Proc. PAC 1995, Particle Accelerator Conference, Dallas, TX, 1995, p. 1642.

[73] R.L. Geng *et al.*, Multipacting in a rectangular waveguide, Proc. PAC 2001, Particle Accelerator Conference, Chicago, IL, 2001, p. 1228.

[74] E. Somersalo *et al.*, Analysis of multipacting in coaxial lines, Proc. PAC 1995, Particle Accelerator Conference, Dallas, TX, 1995, p. 1500.

[75] B. Yunn and R. M. Sundelin, Field emitted electron trajectories for the CEBAF cavity, Proc. PAC1993, Particle Accelerator Conference, Washington, D.C., 1993, p. 1092.

[76] L. Phillips *et al.*, New window design options for CEBAF energy upgrade, Proc. PAC 1997, Particle Accelerator Conference, Vancouver, Canada, 1997, p. 3102.

[77] W.-D. Moeller *et al.*, Development and testing of RF double window input power couplers for TESLA, Proc. 12th Workshop on RF Superconductivity, Ithaca, NY, 2005, p. 571.

[78] S. Noguchi *et al.*, Recent status of the TRISTAN superconducting RF system, Proc. EPAC 1994, European Particle Accelerator Conference, London, 1994, p. 1891.

[79] E. Chojnacki, Tests and Designs of High-Power Waveguide Vacuum Windows at Cornell, Proceedings of the 1997 Workshop on RF Superconductivity, Abano Terme (Padova), Italy, p. 753 (1997).

[80] W.-D. Moeller, High power coupler for the TESLA test facility, Proc. 9th Workshop on RF Superconductivity, Santa Fe, NM, 1999, p. 577.

[81] J. Sekutowicz, Higher order mode coupler for TESLA, TESLA note 1994-07 (1994).

[82] S. Tariq and T. Khabiboulline, FNAL 3.9 GHz HOM coupler & coaxial cable thermal FEA, Proc. 12th Workshop on RF Superconductivity, Ithaca, NY, 2005, p. 604.

[83] E. Harms, Status of 3.9-GHz superconducting RF cavity technology at Fermilab, Proc. LINAC 2006, Linear Accelerator Conference, Knoxville, TN, 2006, p. 695.

[84] G. Wu *et al.*, Electromagnetic simulations of coaxial type HOM coupler, Proc. 12th Workshop on RF Superconductivity, Ithaca, NY, 2005, p. 600.

[85] N. Solyak, New design of the 3.9 GHz HOM coupler, TTC Meeting, KEK, Sept. 25–28, 2006.

[86] C. E. Reece *et al.*, High thermal conductivity cryogenic RF feedthroughs for higher order mode couplers, Proc. PAC 2005, Particle Accelerator Conference, Knoxville, TN, 2005, p. 4108.

[87] I. Campisi, Artificial dielectric ceramics for CEBAF's higher-order mode loads, Proc. 6th Workshop on RF Superconductivity, Newport News, VA, 1993, p. 587.

[88] T. Furuya *et al.*, Superconducting accelerator cavity for KEK B-Factory, Proc. 7th Workshop on RF Superconductivity, Gif-sur-Yvette, France, 1995, p. 729.

[89] S. Belomestnykh *et al.*, Comparison of the predicted and measured loss factor of the superconducting cavity assembly for the CESR upgrade, Proc. PAC 1995, Particle Accelerator Conference, Dallas, TX, 1996, p. 3394.

[90] S. Simrock, Review of slow and fast tuners, Proc. 12th Workshop on RF Superconductivity, Ithaca, NY, 2005, paper ThA07.

[91] E.F. Daly, Overview of existing mechanical tuners, ERL Workshop, 18–23 March, 2005, oral communication.

[92] S. Noguchi, Review of new tuner designs, Proc. 13th Workshop on RF Superconductivity, Beijing, China, 2007, paper WE303.

[93] S. Simrock, Control of microphones and Lorentz force detuning with a fast mechanical tuner, Proc. 11th Workshop on RF Superconductivity, Travemünde, Germany, 2003, paper TuO09.

[94] G. Devanz, Active compensation of Lorentz force detuning of a TTF 9-cell cavity in CRYHOLAB, Proc. LINAC 2006, Linear Accelerator Conference, Knoxville, TN, 2006, p. 598.

[95] M. Liepe, *et al.*, Dynamic Lorentz force compensation with a fast piezoelectric tuner, Proc. PAC 2001, Particle Accelerator Conference, Chicago, IL, 2001, p.1074.

[96] S. Noguchi, Review of new tuner designs, Proc. 13th Workshop on RF Superconductivity, Beijing, China, 2007, paper WE303.

[97] P. Bosland, Tuning systems for superconducting cavities at Saclay, SOLEIL Workshop, 2007, 2007/ESLS-RF/ESLS-RF-PRESENTATIONS/07-ESLS07-PBosland.pdf.